\documentclass[journal]{IEEEtran}

\usepackage{pgfplots}
\usepackage{pgf,tikz,tikz-3dplot}
\usepackage{mwe,tikz}\usepackage[percent]{overpic}
\usetikzlibrary{arrows}
\usepackage{amssymb}
\usepackage{parskip}
\usepackage{graphicx} 
\usepackage{bm}

\usepackage{amsmath,verbatim}
\usepackage{cases}
\usepackage{subfigure}
\usepackage[hide]{ed}

\newcommand{\ee}{\text{e}}
\newcommand{\sd}{\text{d}}          

\def\ro{\vec{r}}
\def\rs{\vec{r}'}
\def\J{\vec{j}}
\def\M{\vec{m}}
\def\E{\vec{e}}
\def\H{\vec{h}}

\def\P{\vec{p}}

\def\Ei{\vec{e}_{\rm inc}}
\def\Hi{\vec{h}_{\rm inc}}
\def\Es{\vec{e}_{\rm sca}}
\def\Hs{\vec{h}_{\rm sca}}

\def\ce{{c_e}}
\def\cm{{c_m}}

\newcommand\Nop{{\cal{N}}}

\newcommand\Kop{{\cal{K}}}

\newcommand{\intV}{\int\limits_{\rm \Omega }}

\newcommand{\ipV}[1]{\langle #1 \rangle_{{\text{\tiny V} }}}

\renewcommand{\vec}[1]{\bm{#1}}

\newcommand{\mat}[1]{\mathbf{#1}}
\newcommand{\matvec}[1]{\mathbf{#1}}

\title{Adiabatic absorbers in photonics simulations with\\
the volume integral equation method\ednote{Title could need some work. Should mention adiabatic absorbers, photonics, and VIE.}}
\author{Alexandra~Tambova, Samuel~P.~Groth, Jacob~K.~White,~\IEEEmembership{Fellow,~IEEE}, \\ and Athanasios~G.~Polimeridis,~\IEEEmembership{Senior Member,~IEEE}

\thanks{Samuel P. Groth and Jacob K. White are with the Department of Electrical Engineering and Computer Science, Massachusetts Institute of Technology, Cambridge, MA 02139, USA.}
\thanks{Alexandra Tambova and Athanasios G. Polimeridis are with the Skolkovo Institute of Technology, Moscow, Russia}}

\begin{document}

\date{\today}
\maketitle

\begin{abstract}
	This paper describes the implementation and performance of adiabatic absorbing layers in an FFT-accelerated volume integral equation (VIE) method for simulating truncated nanophotonics structures. At the truncation sites, we place absorbing regions in which the conductivity is increased gradually in order to minimize reflections. In the continuous setting, such \textit{adiabatic absorbers} have been shown via coupled-mode theory to produce reflections that diminish at a rate related to the smoothness of the absorption profile function. The VIE formulation we employ relies on uniform discretizations of the geometry over which the continuously varying fields and material properties are represented by piecewise constant functions. Such a discretization enables the acceleration of the method via the FFT and, furthermore, the introduction of varying absorption can be performed in a straightforward manner without compromising this speedup. We demonstrate that, in spite of the crude discrete approximation to the smooth absorption profiles, our approach recovers the theoretically predicted reflection behavior of adiabatic absorbers. We thereby show that the FFT-accelerated VIE method is an effective and fast simulation tool for nanophotonics simulations. 

\end{abstract}

\begin{IEEEkeywords}
Integral equations, nanophotonics, adiabatic absorbers, method of moments (MoM), fast solvers.
\end{IEEEkeywords}

\section{Introduction}
In recent years, numerical simulation has become an indispensable tool in the component design process for silicon photonics devices~\cite{chrostowski2015silicon}. Fast and reliable electromagnetics (EM) solvers are used to cheaply prototype new components such as ring resonators and Mach-Zehnder interferometers, and to test their resilience to manufacturing defects such as surface wall roughness~\cite{lee2014low}. The most popular EM solvers in silicon photonics are, at present, those based on approximating Maxwell's equations directly via finite element or finite difference methods.\ednote{should I mention methods such as MEEP, Lumerical, Comsol?} We shall refer to such approaches as differential equation (DE) methods since they discretize the differential operator directly. An alternative approach is to reformulate Maxwell's equations as either surface or volume integral equations over the structure of interest. It is well-known that the integral equation (IE) approach gives rise to a dense matrix system in contrast to the sparse matrices of DE methods. Storing such a dense matrix requires $\mathcal{O}(N^2)$ memory, where $N$ is the number of unknowns. However, fast solvers with $\mathcal{O}(N\log N)$ complexity for IE methods have been developed (e.g.,~\cite{sarkar1986application,sertel2004multilevel}), thus allowing them to be competitive with DE methods. Furthermore, IE methods have the distinct advantage that they are dispersion free owing to the fact that the Green's function is an exact propagator of the field~\cite{chew2008integral}. This dispersion-free property is especially desirable in the nanophotonics setting where the structures of interest may span hundreds or thousands of wavelengths over which dispersion could potentially lead to large phase errors.

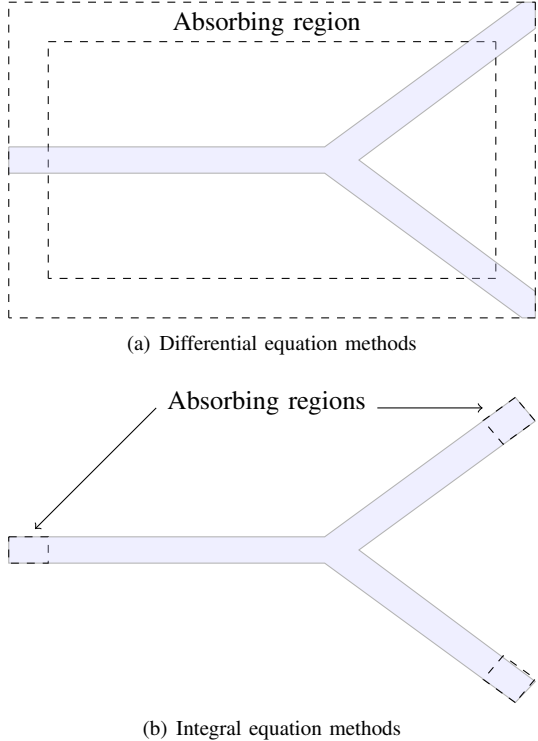
\begin{figure}[h!]
\centering
\subfigure[Differential equation methods]{
\begin{tikzpicture}[scale=0.7]
	\draw[dashed] (-5,3) -- (5,3);
	\draw[dashed] (5,3) -- (5,-3);
	\draw[dashed] (5,-3) -- (-5,-3);
	\draw[dashed] (-5,-3) -- (-5,3);
	
	\draw[dashed] (-4.25,2.25) -- (4.25,2.25);
	\draw[dashed] (4.25,2.25) -- (4.25,-2.25);
	\draw[dashed] (4.25,-2.25) -- (-4.25,-2.25);
	\draw[dashed] (-4.25,-2.25) -- (-4.25,2.25);
	
	\filldraw[fill=blue!20,opacity=0.3] (-5,0.25) -- (1,0.25) -- (4.75,3) -- (5,3) -- (5,2.5)
	           -- (1.65,0) -- (5,-2.5)  -- (5,-3) -- (4.75,-3) -- (1,-0.25) -- (-5,-0.25) -- (-5,0.25);
	           
	\draw (-0.1,3) node[below]{Absorbing region}; 
\end{tikzpicture}
}
\hspace{1cm}
\subfigure[Integral equation methods]{
\begin{tikzpicture}[scale=0.7]
	
	\draw[dashed] (-5,0.25)--(-4.25,0.25)--(-4.25,-0.25)--(-5,-0.25)--(-5,0.25);
	
	\filldraw[fill=blue!20,opacity=0.3] (-5,0.25) -- (1,0.25) -- (4.62,2.9) -- (5.0,2.45)
	           -- (1.65,0) -- (5,-2.5)  -- (4.62,-2.9) -- (1,-0.25) -- (-5,-0.25) -- (-5,0.25);
	           
	 \draw[dashed] (4.62,2.9) -- (4.01,2.46) -- (4.39,2.01)  -- (5,2.45) -- (4.62,2.9) ;
	 
	  \draw[dashed] (4.62,-2.9) -- (4.01,-2.46) -- (4.39,-2.01)  -- (5,-2.45) -- (4.62,-2.9) ;
	           
	\draw (-0.1,3.2) node[below]{Absorbing regions}; 
	
	\draw[->] (2,2.7) -- (4,2.7);
	\draw[->] (-2.2,2.7) -- (-4.5,0.4);
	
\end{tikzpicture}
}

\caption{A 2D-slice comparison of use of absorbing layers/PMLs in differential equation (FE/FD) methods and integral equation (VIE/SIE) methods.}
\label{fig:splitter_PML}
\end{figure}

Integral equation techniques have traditionally found their application in exterior scattering problems where a wave impinges on a finite obstacle and is scattered into the surrounding infinite volume. IEs are desirable in this context since they satisfy the radiation condition at infinity by construction and reduce a computation over the infinite scattering domain to one merely over the finite obstacle. In contrast, a DE method would have to truncate the infinite domain at some distance from the obstacle with an absorbing region or perfectly matched layer (PML). When applied to unbounded obstacles such as waveguides, IEs can be applied with a modified Green's function constructed especially to take into account the unbounded nature of the particular geometry. However, this is an involved approach and so far has been applied successfully only for 2D waveguide-type problems, e.g., in~\cite{hanson2003green,kamandi2015integral}. A more pragmatic and flexible approach is to introduce absorbing regions, as is done in DE methods. One advantage of an IE approach over DE is that the deployment of absorbing regions can be more flexible and, further, they are required over a much smaller region. Consider Fig.~\ref{fig:splitter_PML} in which we compare the use of absorbing regions in DE and IE methods for a simple waveguide splitter taken from \cite{oskooi2008failure}. In DE methods, one must artificially truncate the entire computational domain, whereas in IE methods, it is only necessary to truncate the portions of the obstacle which extend away in an unbounded fashion; here it is the waveguide branches or ports. The remainder of the domain is truncated analytically via the IE formulation. This leads to a considerably reduced computation domain. Further note that the diagram in Fig.~\ref{fig:splitter_PML} is of a 2D slice. In DE methods there must also be absorbing layers above and below the waveguide whereas the IE method truncates in these directions analytically by construction.

For domain truncations, PMLs have previously been used in IE settings~\cite{alles2011perfectly}, however it is known that for certain scenarios, PMLs fail~\cite{oskooi2008failure}. In particular, when the material properties are not analytic functions in the direction perpendicular to the PML boundary. We consider one such example (a Bragg grating) in detail in Section~\ref{ss:slow_light}. It is shown in \cite{oskooi2008failure} (albeit there in the context of FD methods) that a more robust approach to domain truncation is to use adiabatic layers rather than PMLs. Adiabatic layers are regions in which the conductivity (absorption) of the medium is gradually increased. It was shown in \cite{zhang2011novel} that adiabatic absorbers are effective for surface integral equations (SIEs). In this paper, we discuss the first implementation of adiabatic absorbers in the \textit{volume} integral equation method. In particular, we use the fast open-source VIE package MARIE~\cite{polimeridis2014stable,MARIE_GIT} for all simulations.

The layout of the paper is as follows. In Section~\ref{sec:setup} the general setup of photonics simulations with the VIE is outlined. This includes a description of the different geometries considered in the numerical results section Section~\ref{sec:num_results}, how the absorbers are appended to the waveguides, and how the waveguides are excited. Section~\ref{sec:VIE} gives a brief review of the VIE formulation and the discretization of the resulting VIEs. Section~\ref{sec:adiabatic} provides details of the adiabatic absorbers and summarizes results pertaining to the reflections they produce as a function of their length and the material properties. In Section~\ref{sec:num_results} we examine three examples: a straight dielectric strip waveguide, a Bragg grating, and a Y-branch splitter. We observe that the adiabatic absorber performs extremely well in the VIE setting, and reproduces the theoretical results from the literature. Finally, in Section~\ref{sec:conc} we provide concluding remarks and discuss briefly some numerical aspects of the performance of the VIE method for these photonics problems. Finally, we discuss potential future improvements in the application of the VIE method to photonics simulations.

\section{Setup for photonics simulations}
\label{sec:setup}
In this paper, we consider three different nanophotonics structures: a dielectric strip waveguide (see Fig.~\ref{fig:strip}), a Bragg grating waveguide similar to that in \cite{povinelli2005slow} (see Fig.~\ref{fig:Bragg}), and a Y-branch splitter. All structures considered consist of a silicon (Si) core surrounded by silicon dioxide (SiO$_2$). The relative permittivity of Si is wavelength dependent and we assume it obeys the Lorentz model~\cite{chrostowski2015silicon,oughstun2003lorentz}, and the relative permittivity of SiO$_2$ is taken to be $1.444^2$~\cite{chrostowski2015silicon}. Note that in Section \ref{sec:VIE}, where the VIE method is described, we assume the exterior medium has unit relative permittivity. In order to make this equivalent to our physical problem, we must scale the relative permittivity of Si and the wavelength of the incident field accordingly (dividing by $1.444^2$ and 1.444, respectively).

In order to excite the waveguide, a dipole is placed on the center-line of the waveguide, $\lambda/4$ from the left end, where $\lambda$ represents the wavelength inside silicon. We add a small imaginary shift to the dipole's location in order to produce a Gaussian beam propagating in the $+x$-direction. The shift we use is $-\lambda j$ which was found to give a good compromise between directionality and localization (for details see \cite{zhang2011novel}).\ednote{The thesis might be more appropriate to reference here.}

\subsection{Dielectric strip waveguide}
\label{ss:strip_geometry}
\begin{figure}[h!]
\centering
\tdplotsetmaincoords{60}{325}
\begin{tikzpicture}[
		scale=0.65,
		tdplot_main_coords,
		axis/.style={->,black,thick},
		cube/.style={very thin,opacity=0.2,fill=red},
		cube hidden/.style={very thin,opacity=.3,gray}]

	\draw[axis] (0,0,0) -- (12.5,0,0) node[anchor=west]{$x$};
	\draw[axis] (0,0,0) -- (0,4,0) node[anchor=east]{$y$};
	
	\draw[cube] (0,0,0) -- (9,0,0) -- (9,0,1) -- (0,0,1) -- cycle;

	\draw[cube] (0,0,0) -- (0,0,1) -- (0,2,1) -- (0,2,0) -- cycle;

	\draw[cube] (0,0,1) -- (0,2,1) -- (9,2,1) -- (9,0,1) -- cycle;
	
	\draw[cube hidden] (9,0,0) -- (9,2,0);
	\draw[cube hidden] (9,2,0) -- (0,2,0);
	\draw[cube hidden] (9,2,0) -- (9,2,1);
	
	\draw[axis] (0,0,0) -- (0,0,3) node[anchor=west]{$z$};

	\foreach \x in {0,0.01,...,2}
		{
 			 \draw[draw=red,opacity=0.1+ 0.22*\x^2]  (9+\x,0,1) -- (9+\x,2,1);
		}

	\foreach \x in {0,0.01,...,2}
		{
 				 \draw[draw=red,opacity=0.1+ 0.22*\x^2]  (9+\x,0,0) -- (9+\x,0,1);
		}
  
	\draw[cube hidden] (11,0,0) -- (11,2,0);
	\draw[cube hidden] (11,2,0) -- (9,2,0);
	\draw[cube hidden] (11,2,0) -- (11,2,1);
	
	\draw (10.1,-0.25,-0.1) node[below]{Absorber};
	
	\draw (5,1.9,0.6) node[below]{Si};
	
	\draw (5,-1,0) node[below]{SiO$_2$};
	
\end{tikzpicture}
\caption{Problem setup for a Gaussian beam source within a silicon strip waveguide with one absorber attached. Only one absorber is necessary for this problem since the source has directionality and we anticipate no reflections propagating in the $-x$-direction. The cladding medium is silicon dioxide.}
\label{fig:strip}
\end{figure}
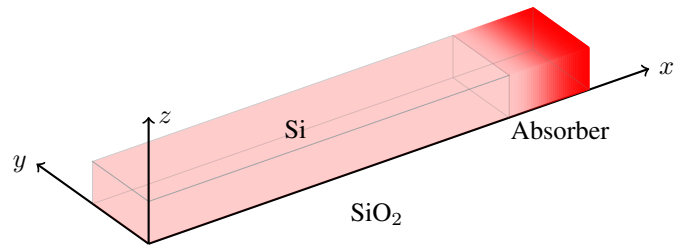

The waveguide we consider in Section~\ref{ss:strip} is depicted in Fig.~\ref{fig:strip}. It occupies the space
\[
	0\leq x\leq 13500\text{nm},\ 0\leq y\leq 500\text{nm},\ 0\leq z\leq 200\text{nm}.
\]
This size, or similar, for the $(y,z)$ cross-section is a popular choice owing to its support of one dominant TE guided mode in the free-space wavelength range of 1500nm to 1600nm. The free-space wavelength of light considered is 1550nm which equates approximately to a wavelength of $\lambda=446$nm within the silicon core. Therefore, the length of the waveguide is roughly 30 wavelengths within silicon. 
In this problem, we expect all the waves to propagate in the $+x$-direction, hence we append only one absorber, on the right-hand end of the waveguide.

To discretize the waveguide, we use voxels of size 50/3nm since this perfectly divides the dimensions of the structure. Further, this represents approximately 27 voxels per wavelength (inside silicon) which is a high enough resolution to ensure accurate simulations.
 
\subsection{Bragg grating}
\label{ss:Bragg_geometry}
The Bragg grating is a fundamental photonics component for filtering out particular wavelengths from a signal~\cite{chrostowski2015silicon}. Along the propagation direction, the grating's width has a periodic modulation. This modulation leads to distributed reflections which only interfere constructively in a narrow wavelength band centered around the \textit{Bragg wavelength}. In this band, the input signal is strongly reflected, resulting in reduced transmission through the grating. We perform a set of simulations in Section~\ref{ss:slow_light} to show this band gap.

A typical Bragg grating is depicted in Fig.~\ref{fig:Bragg}. Here we shall use the following values for the geometrical parameters in the figure:
\begin{align*}
&	D=220\text{nm},\ W=500\text{nm},\ \Delta W=40\text{nm},\\
& \Lambda=320\text{nm},\ N=100.
\end{align*}
\def\nn{15}
\begin{figure}[ht!]
\centering
\begin{tikzpicture}[scale=0.3]

\shade[left color=white,right color=black] (\nn+7,0) rectangle (\nn+9,2);
\shade[left color=black,right color=white] (-4,2) rectangle (-2,0);

\draw (1,2) -- (1,2.25) -- (2,2.25) -- (2,1.75);
\draw (1,0) -- (1,-0.25)-- (2,-0.25)-- (2,0.25);

\foreach \i in {1,3,5,7,9,11,13,\nn}
{
	\draw (\i+1,1.75)--(\i+2,1.75)--(\i+2,2.25)--(\i+3,2.25)--(\i+3,1.75);
	\draw (\i+1,0.25)--(\i+2,0.25)--(\i+2,-0.25)--(\i+3,-0.25)--(\i+3,0.25);
}

\draw (\nn+3,1.75) -- (\nn+4,1.75) -- (\nn+4,2) ;
\draw (\nn+3,0.25) -- (\nn+4,0.25) -- (\nn+4,0) ;

\draw (\nn+4,0) -- (\nn+9,0) -- (\nn+9,2) -- (\nn+4,2);
\draw (1,0) -- (-4,0) -- (-4,2) -- (1,2);

\draw [|-|] (5,2.75) -- (7,2.75);
\draw (6,2.9) node[anchor=south]{$\Lambda$};

\draw [|-|] (1,-0.75) -- (\nn+4,-0.75);
\draw (10,-1) node[anchor=north]{$N\times\Lambda$};

\draw [->] (12.5,1.0)--(12.5,1.7);
\draw [->] (12.5,3.0)--(12.5,2.3);
\draw (12.5,2.8) node[anchor=west]{$\Delta W$};

\draw [<->] (-0.75,0.1)--(-0.75,1.9);
\draw (-0.75,1) node[anchor=west]{$W$};

\draw[->] (\nn+8,-1)node[anchor=north]{Absorber}--(\nn+8,-0.1);

\end{tikzpicture}

\caption{Top view of the layout for Bragg grating with period $\Lambda$, width $W$, corrugation depth $\Delta W$, and length $N\times\Lambda$, where $N$ is an integer. Since the waves propagate in both directions due to reflections from the corrugations, absorbers are required on both ends for simulations. To generate the 3D structure, this layout is extruded a distance $D$ in the $z$-direction (out of the page).}
\label{fig:Bragg}
\end{figure}
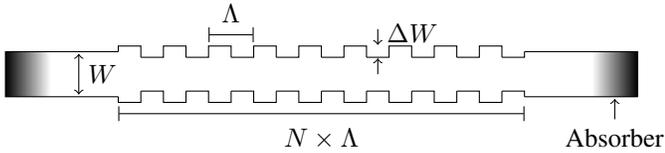
The periodic part of the structure is preceded by a uniform region of length $10\Lambda\ (\approx7\lambda)$, which is excited using the dipole located in the same position as in the previous setup for the strip waveguide. The characteristic modulation in the waveguide leads to reflections and hence waves propagate in both directions. Therefore, we require absorbing regions on both ends to truncate the structure, in this example.

To discretize this geometry, and the Y-branch splitter to follow, we use voxels of size 20nm. This size voxel is chosen since it can perfectly represent the cross section of the waveguides. Furthermore, at approximately 22 voxels per interior wavelength, this resolution is fine enough to ensure accurate simulations. 

\subsection{Y-branch splitter}
\label{ss:y_branch_geom}
\begin{figure}[ht!]
	\includegraphics[width=0.45\textwidth]{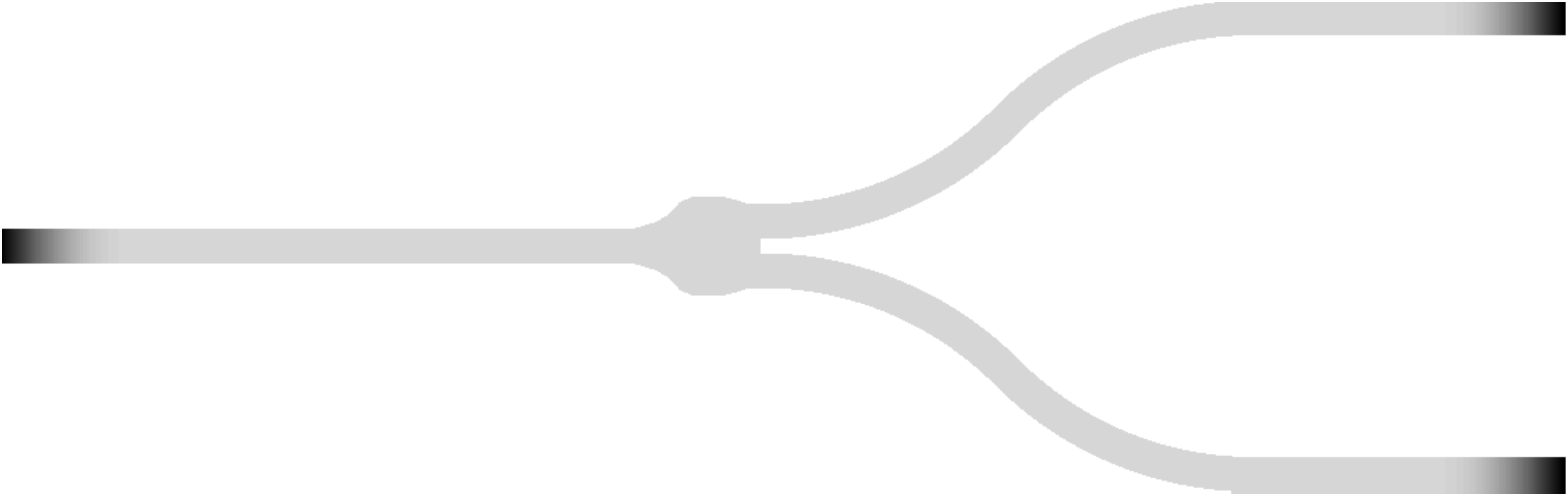}
	\caption{Top view of the layout of a Y-branch splitter with adiabatic absorbers appended to the waveguide-branch ends. This structure is excited by a Gaussian beam injected just to the right of the left absorber. A right-propagating mode is established, is split in half at the junction, and each half propagates along its respective curved branch. The geometry for the junction is taken from \cite{zhang2013compact}.}
	\label{fig:splitter}
\end{figure}

The final geometry we consider is the Y-branch splitter/combiner shown in Fig.~\ref{fig:splitter}. The role of this device is to split a signal from one waveguide equally into two, or to combine two separate signals into one. The particular geometry we use is similar to that in \cite{chrostowski2015silicon,zhang2013compact}. The individual waveguide branches each have $(x,y)$-cross section dimensions $(W,D)$:
\[
	W = 500\text{nm},\quad D=220\text{nm},
\]
and the geometry of the junction is described in detail in \cite{zhang2013compact}.

We terminate the ends of the waveguides using adiabatic absorbers. The absorber lengths and absorption profiles shall be specified in the results section.

\section{Volume integral equations}
\label{sec:VIE}
We briefly summarize the VIE formulation we solve in this paper. For a detailed derivation, see, e.g., \cite{polimeridis2014stable,volakis2012integral}.
 
We consider the scattering\ednote{could say ``propagation'' instead to avoid association with the scattering community} of time-harmonic electromagnetic waves with angular frequency $\omega$ by\ednote{or ``through''} a dielectric, potentially inhomogeneous, object occupying a bounded domain $\Omega$ in 3D space $\mathbb{R}^3$. Throughout the time-dependance $\ee^{j\omega t}$ is assumed with $j=\sqrt{-1}$. The electric and magnetic properties are defined as
\begin{equation}
\begin{split}
	& \epsilon = \epsilon_0,\ \mu = \mu_0 \quad \text{in}\ \mathbb{R}^3\backslash\Omega, \\
	& \epsilon = \epsilon_{r}(\ro)\epsilon_0,\ \mu=\mu_{r}(\ro)\mu_0 \quad \text{in}\ \Omega, 
\end{split}
\end{equation}
where $\epsilon_0$ and $\mu_0$ are the free-space permittivity and permeability, respectively. The relative permittivity and permeability are written
\begin{equation}
\begin{split}
	\epsilon_r(\ro) & = \epsilon_r'(\ro) - j\epsilon_r''(\ro), \\
	\mu_r(\ro)         & = \mu_r'(\ro) - j\mu_r''(\ro),
\end{split}
\end{equation}
with $\epsilon_r',\mu_r'\in(0,\infty)$ and $\epsilon_r'',\mu_r''\in[0,\infty)$.

The total electric and magnetic fields $(\E,\H)$ are composed of incident and scattered fields
\begin{equation}
	\left(\begin{array}{c}
		\E \\
		\H
	\end{array}\right)
	=
	\left(\begin{array}{c}
		\Ei \\
		\Hi
	\end{array}\right)
	+
	\left(\begin{array}{c}
		\Es \\
		\Hs 
	\end{array}\right),
	\label{eqn:field_decomp}
\end{equation}
where the incident fields $(\Ei,\Hi)$ are generated by dipoles or Gaussian beams in the absence of the scatterer. The scattered fields can be expressed in terms of equivalent polarization and magnetization currents $(\J,\M)$ as
\begin{equation}
\begin{split}
	\left(\begin{array}{c}
		\Es \\
		\Hs
	\end{array}\right)
	&=
	\left(\begin{array}{cc}
		\frac{1}{c_e}(\mathcal{N}-\mathcal{I}) & -\mathcal{K} \\
		\mathcal{K} & \frac{1}{c_m}(\mathcal{N}-\mathcal{I})
	\end{array}\right)
	\left(\begin{array}{c}
		\J\\
		\M
	\end{array}\right),
	\label{eqn:int_rep}
\end{split}
\end{equation}
where $c_e:=j\omega\epsilon_0,\ c_m:=j\omega\mu_0$, and $\mathcal{I}$ is the identity operator. The integro-differential operators are defined as
\begin{align}
	\mathcal{K}f &:= \nabla\times\mathcal{S}(f), \\
	\mathcal{N}f &:= \nabla\times\nabla\times\mathcal{S}(f),
\end{align}
where
\begin{equation}
	\mathcal{S}(f) := \int_{\Omega} G(\ro-\rs)f(\rs)\sd \rs
\end{equation}
is the volume vector potential, $G$ is the free-space scalar Green's function
\begin{equation}
	G(\ro) := \frac{\ee^{-j k_0|\ro-\rs|}}{4\pi |\ro-\rs|}, 
\end{equation}
and $k_0=\omega\sqrt{\epsilon_0\mu_0}$ is the free-space wavenumber. The equivalent current densities are given in terms of the fields as
\begin{equation}
\begin{split}
	\J(\ro) &= c_e(\epsilon_r(\ro)-1)\E(\ro),\\
	\M(\ro) &= c_m(\mu(\ro)-1)\H(\ro).
\end{split}
\label{eqn:equiv_currents}
\end{equation}

The JM-VIE formulation can be derived by combining (\ref{eqn:field_decomp}), (\ref{eqn:int_rep}) and (\ref{eqn:equiv_currents}) to obtain (see \cite{polimeridis2014stable,markkanen2012discretization} for more details):
\begin{equation}
\boxed{
	\left(\boldsymbol{\mathcal{I}} - \boldsymbol{\mathcal{M}}\boldsymbol{\mathcal{T}}\right)
	\left(\begin{array}{c}
		\J \\
		\M 
	\end{array}\right)
	= \boldsymbol{\mathcal{C}}\boldsymbol{\mathcal{M}}
	\left(\begin{array}{c}
		\Ei \\
		\Hi
	\end{array}\right),}
	\label{eqn:JM_VIE}
\end{equation}
where 
\begin{equation}
	\boldsymbol{\mathcal{T}}=
	\left(\begin{array}{cc}
		\mathcal{N} & -c_e\mathcal{K} \\
		c_m\mathcal{K} & \mathcal{N}
	\end{array}\right),
\end{equation}
and
\begin{equation}
	\boldsymbol{\mathcal{M}}=
	\left(\begin{array}{cc}
		\mathcal{M}_{\epsilon}  & 0 \\
		0 & \mathcal{M}_{\mu}
	\end{array}\right),
	\quad
	\boldsymbol{\mathcal{C}}=
	\left(\begin{array}{cc}
		c_e\mathcal{I}  & 0 \\
		0 & c_m\mathcal{I}
	\end{array}\right).
\end{equation}
Here, $\mathcal{M}_{\epsilon}$ and $\mathcal{M}_{\mu}$ are multiplication operators that multiply by the respective local functions $(\epsilon_r(\ro)-1)/\epsilon_r(\ro)$ and $(\mu_r(\ro)-1)/\mu_r(\ro)$.

The full JM-VIE formulation (\ref{eqn:JM_VIE}) is not necessary for the photonics applications of interest here since magnetic currents are not present. However, we include this formulation since we will later consider the effect on absorber quality of introducing magnetic conductivity alongside electric conductivity. For the majority of the paper we shall instead use the J-VIE formulation which is simply obtained from (\ref{eqn:JM_VIE}) by setting the magnetic current densities, $\M$, to zero, giving
\begin{equation}
\boxed{
	\left(\mathcal{I} - \mathcal{M}_{\epsilon}\mathcal{N}\right)\J = c_e\mathcal{M}_{\epsilon}\Ei.}
	\label{eqn:JVIE}
\end{equation}

Observe that the integral operators in the formulations (\ref{eqn:JM_VIE}) and (\ref{eqn:JVIE}) are both of the form \textit{identity plus diagonal multiplier times compact}. Such operators are desirable in our setting for two main reasons: firstly, they are \textit{second kind} integral operators which are well behaved in terms of accuracy and convergence; secondly, the influence of the material properties is confined to the diagonal multiplier $\boldsymbol{\mathcal{M}}$. This second point means that the implementation of absorbing regions is particularly simple in this VIE setting, since all we have to do is alter the entries in the multiplier $\boldsymbol{\mathcal{M}}$ in order to introduce absorption, with the rest of the machinery remaining unchanged. Furthermore, as we discuss in the next section, the discrete forms of $\mathcal{N}$ and $\mathcal{K}$ both have Toeplitz structure when uniform meshing is employed, which enables the FFT-acceleration of the VIE method. This desirable structure is unaffected by perturbing $\boldsymbol{\mathcal{M}}$, hence the fast nature of the method remains.

\subsection{Discretization}
There are numerous discretization techniques available for numerically solving the JM-VIE (\ref{eqn:JM_VIE}). Here we employ the Galerkin method over a uniform (``voxelized'') discretization of the domain. We represent the unknown currents $\J,\M\in[L^2(\mathbb{R}^3)]^3$ as piecewise constant functions on this voxelized grid:
\begin{equation}\label{JM_exp}
\J \approx \sum\limits_{i} {w_e}_i \P_i,\quad
\M \approx \sum\limits_{i} {w_m}_i \P_i,
\end{equation}
where the weights ${w_e}_i, {w_m}_i$ are to be determined, and 
\begin{equation}
	\P_i = \frac{1}{(\Delta V)^{1/2}}
\end{equation}
is a constant function with support restricted to voxel $V_i$. The scaling the square root of the voxel volume $\Delta V$ is included so that
\begin{equation}
	\langle \P_i,\P_j \rangle = \delta_{ij},
\end{equation}
where $\langle\cdot,\cdot\rangle$ is the standard $L^2$ inner product and $\delta_{ij}$ is the Kronecker delta. 

Applying the Galerkin method to the JM-VIE (\ref{eqn:JM_VIE}), with testing functions $\P_i$, gives rise to the linear system
\begin{equation}
	\left(\mat{I} - \mat{A}\right)
	\left(\begin{array}{c}
		\matvec{w}_e     \\
		\matvec{w}_m
	\end{array}\right)
	= \mat{C}
	\left(\begin{array}{c}
		\matvec{b}_{\text{e}} \\
		\matvec{b}_{\text{e}}
	\end{array}\right),
	\label{eqn:JM_VIE_discrete_general}
\end{equation}
where $\mat{I}$ is the identity matrix; the discrete form of the integral operator is
\begin{equation}
	\mat{A}=
	\left(\begin{array}{cc}
		\mat{A}_{11}  & -c_e\mat{A}_{12} \\
		c_m\mat{A}_{21} & \mat{A}_{22}
	\end{array}\right), \\
\end{equation}
where
\begin{align}
&	(\mat{A}_{11})_{ij} = \langle \mathcal{M}_{\epsilon}\mathcal{N}\P_j,\P_i\rangle,\quad  (\mat{A}_{12})_{ij} = \langle \mathcal{M}_{\epsilon}\mathcal{K}\P_j,\P_i\rangle, \\
& 	(\mat{A}_{21})_{ij} = \langle \mathcal{M}_{\mu}\mathcal{K}\P_j,\P_i\rangle,\quad ( \mat{A}_{22})_{ij} = \langle \mathcal{M}_{\mu}\mathcal{N}\P_j,\P_i\rangle;
\end{align}
and the right-hand side is
\begin{equation}
	(\mat{b}_{\text{e}})_i = \langle \mathcal{M}_{\epsilon}\Ei,\P_i\rangle,\quad  (\mat{b}_{\text{h}})_i = \langle \mathcal{M}_{\mu}\Hi,\P_i\rangle.
\end{equation}

In this paper, we represent the material properties, encapsulated in $\mathcal{M}_{\epsilon}$ and $\mathcal{M}_{\mu}$, as piecewise constant functions across the voxel grid. That is, we assume that $\mathcal{M}_{\epsilon}$ and $\mathcal{M}_{\mu}$ are constant on each voxel with its value being defined at the voxel centers. This enables $\mathcal{M}_{\epsilon}$ and $\mathcal{M}_{\mu}$ to be removed outside the inner products above, thereby allowing (\ref{eqn:JM_VIE_discrete_general}) to be written in the following simplified form:
\begin{equation}
\boxed{
	\left(\mat{I} - \mat{M}\mat{T}\right)
	\left(\begin{array}{c}
		\matvec{w}_e     \\
		\matvec{w}_m
	\end{array}\right)
	= \mat{C}\mat{M}
	\left(\begin{array}{c}
		\matvec{e}_{\text{inc}} \\
		\matvec{h}_{\text{inc}}
	\end{array}\right),}
	\label{eqn:JM_VIE_discrete}
\end{equation}
where $\mat{I}$ is the identity matrix; the diagonal material properties multipliers are
\begin{equation}
	\mat{M}=
	\left(\begin{array}{cc}
		\mat{M}_{\epsilon}  & 0 \\
		0 & \mat{M}_{\mu}
	\end{array}\right),
	\quad
	\mat{C}=
	\left(\begin{array}{cc}
		c_e\mat{I}  & 0 \\
		0 & c_m\mat{I}
	\end{array}\right); 
\end{equation}
the discrete form of the integral operator is
\begin{equation}
	\mat{T}=
	\left(\begin{array}{cc}
		\mat{N}  & -c_e\mat{K} \\
		c_m\mat{K} & \mat{N}
	\end{array}\right), \\
\end{equation}
where
\begin{equation}
	\mat{N}_{ij} = \langle \mathcal{N}\P_j,\P_i\rangle,\quad  \mat{K}_{ij} = \langle \mathcal{K}\P_j,\P_i\rangle;
\end{equation}
and the right-hand side is
\begin{equation}
	(\mat{e}_{\text{inc}})_i = \langle \Ei,\P_i\rangle,\quad  (\mat{h}_{\text{inc}})_i = \langle \Hi,\P_i\rangle.
\end{equation}

The uniform discretization we use is desirable because it results in the matrices $\mat{N}$ and $\mat{K}$ being Toeplitz, hence matrix-vector products using $\mat{N}$ and $\mat{K}$ can be performed in $\mathcal{O}(N\log N)$ operations with the use of the FFT, where $N$ is the number of voxels. Further, the piecewise constant representation of the material properties means that introducing varying conductivity in an absorbing region does not interfere with the Toeplitz structure of $\mat{N}$ and $\mat{K}$. It only affects the diagonal entries in the multiplier $\mat{M}$. This makes the implementation of absorbing regions in the VIE method particularly straightforward, and does not compromise the FFT-acceleration. We see in Section~\ref{sec:num_results} that, even with this crude piecewise constant representation for higher-order polynomial conductivity profiles, our approach still recovers the asymptotic behavior of the continuous analogues of these profiles.

\section{Reflections from adiabatic absorbers}
\label{sec:adiabatic}
\subsection{Generic adiabatic absorbers}
As mentioned in the introduction, an adiabatic absorber is a region in which absorption is turned on gradually in order to reduce reflections at the absorber interface. Specifically, we define the absorption profile as
\begin{equation}
	\sigma(x) =
	\begin{cases}
		0, & x<0, \\
		 \sigma_0 s(x/L), & 0\leq x\leq L,
	\end{cases}
	\label{eqn:abs_profile}
\end{equation}
where $x=0$ is the beginning of the absorber of length $L$. Observe that we have introduced a scaled coordinate $u=x/L\in [0,1]$ for ease of presentation later on. Note further that $\sigma$ can represent either of the electric or magnetic conductivities, $\sigma_e, \sigma_m$. In this paper, we consider the first four monomials as our candidate absorption profiles:
\begin{equation}
	s(u) =
	\begin{cases}
		0, & u<0, \\
		 u^d, & 0\leq u\leq 1,
	\end{cases}
	\label{eqn:profiles}
\end{equation}
for $d=0,1,2,3$. During this discussion of reflections from absorbers, we shall refer specifically to these monomial absorption profiles.

\subsection{Adiabatic absorbers in the EM setting}
Here we discuss the form of our adiabatic absorbers in two cases: when we include both electric and magnetic conductivity, and when we include only electric conductivity. As we shall see, the former allows for impedance matching and hence superior absorbers, but at the cost of solving for twice as many unknowns.

Consider the simple waveguide setup depicted in Fig.~\ref{fig:strip}. Suppose that the waveguide begins at the origin and extends to $x=X$ before the absorbing region begins, and this region terminates at $x=X+L$. Assume that the permittivity for $0\leq x\leq X+L$ has the form:
\begin{equation}
	\epsilon_r(x) =
	\begin{cases}
		\epsilon_r'(x), &\quad 0\leq x\leq X, \\
		 \epsilon_r'(x) - j\epsilon_r''(x), & \quad X< x\leq X+L.
	\end{cases}
	\label{eqn:permittivity}
\end{equation} 
That is, the permittivity before the absorber is real and in the absorber is complex.

\textit{If we are including magnetic conductivity}, we have that the magnetic permeability is
\begin{equation}
	\mu_r(x) =
	\begin{cases}
		1, &\quad 0\leq x\leq X, \\
		1 - j\mu_r''(x), & \quad X< x\leq X+L.
	\end{cases}
	\label{eqn:permability}
\end{equation} 
In order to match the impedances between the waveguide and the absorbing region, we set
\begin{equation}
	\mu_r''(x) = \frac{\epsilon_r''(x)}{\epsilon_r'(x)}\quad\text{for}\ X< x\leq X+L.
\end{equation}
However, \textit{if we are not including magnetic conductivity}, we have that
\begin{equation}
	\mu_r(x) = 1,\quad \text{for all}\ x.
\end{equation}
In this case, we may simply set $\M=0$ in (\ref{eqn:JM_VIE}), thereby halving the number of unknowns.

\subsection{Round-trip reflection and transition reflection}
We consider two types of reflection caused by the adiabatic layer, namely the round-trip and transition reflections. The round-trip reflection, $R_{\text{rt}}$, is the reflection due to the wave propagating all the way to the end of the absorber, reflecting off the end, and returning back. Whereas the transition reflection, $R_{\text{t}}$, is the reflection of the wave at the absorber's interface. We can derive approximate expressions for these two types of reflection. For the round-trip reflection, we can just consider the exponentially decaying wave as it propagates to the end of the absorber and back. For the transition reflection, we appeal to results from coupled-mode theory~\cite{johnson2002adiabatic}.

\begin{center}
	\textbf{Round-trip reflection}
\end{center}
The round-trip reflection $R_{\text{rt}}$ can be shown to take the form
\begin{equation}
	R_{\text{rt}}\sim \exp\left\{-D\eta_x k_0\int_0^L\frac{\epsilon_r''(x)}{\sqrt{\epsilon_r'(x)}}\sd x\right\},
	\label{eqn:decay}
\end{equation}
where $0\leq\eta_x\leq 1$. For a plane wave propagating purely in the $x$-direction, $\eta_x=1$. When the impedance is matched, we have that $D=4$; this comes from the fact that the wave travels a distance $2L$, then this is squared to obtain the reflected power. When the impedance is not matched, it can be shown that $D=2$. The factor of two difference can be attributed to the presence of two attenuating mechanisms in the matched impedance case, namely the decay due to both the magnetic and electric conductivities, whereas there is only the electric conductivity in the unmatched case.

Writing $\epsilon_r''(x)/\sqrt{\epsilon_r'(x)}=\sigma_0 s(x/L)/\left(\omega\epsilon_0\right)$ and making the change of variables $u=x/L$ leads to the following form of (\ref{eqn:decay}):
\begin{equation}
	R_{\text{rt}}\sim \exp\left\{-D\eta_x\sqrt{\frac{\mu_0}{\epsilon_0\epsilon_{r}}}L\sigma_0\int_0^1 s(u)\sd u\right\},
	\label{eqn:decay_2}
\end{equation}	
where we have also made use of the identity $k_0=\omega \sqrt{\mu_0\epsilon_0}$. For the monomial absorption profiles (\ref{eqn:profiles}) under consideration in this paper, this simplifies to
\begin{equation}
	R_{\text{rt}}\sim \exp\left\{-\frac{DL\eta_x\sqrt{\frac{\mu_0}{\epsilon_0\epsilon_{r}}}\sigma_0}{p+1}\right\}.
	\label{eqn:rt_asymp}
\end{equation}
Now we can choose $\sigma_0$ such that we obtain a desired round-trip reflection $R_{\text{rt}}$ via the following formula
\begin{equation}
	\sigma_0 = -\frac{(p+1)\ln(R_{rt})}{DL\eta_x}\sqrt{\frac{\epsilon_0 \epsilon_{r}}{\mu_0}}.
	\label{eqn:sigma0}
\end{equation}
In the experiments section, we shall set $\eta_x=1$ which may lead to a slight under-estimation of the round-trip reflection magnitude. (For a propagating mode, we expect $\eta_x<1$ to be the ratio of the propagation constant to the interior wavenumber, but we do not in general know $\eta_x$ a priori.) We choose values of $\sigma_0$ high enough such that we can be certain that the round-trip reflection is much smaller than the error in our numerical scheme, even in the presence of this under-estimation.

\begin{center}
	\textbf{Transition reflection}
\end{center}
An effective way to analyze the propagation of waves along a waveguide with slowly varying properties (in this case, the conductivity) is via \textit{coupled-mode theory} (CMT). Here we quote the appropriate results from \cite{johnson2002adiabatic} where the details can be found in full. For a concise summary of the pertinent details of CMT and the results of \cite{johnson2002adiabatic}, the reader is referred to \cite{oskooi2008failure}. It is shown in \cite{johnson2002adiabatic} that in the limit of slow variation in conductivity (equivalently, the large $L$ limit), the amplitude $c_r$ of a reflected mode has the asymptotic form
\begin{equation}
	c_r(L) = s^{(d)}(0^+)\frac{M(0^+)}{\Delta\beta(0^+)}[-j L\Delta\beta(0^+)]^{-d}+\mathcal{O}(L^{-(d+1)}),
	\label{eqn:cr}
\end{equation}
where $s^{(d)}(0^+)$ is the first non-zero derivative of the absorption profile $s(u)$ at $u=\frac{x - X}{L} = 0^+$. Here $M$ is a coupling coefficient between the incident and reflected modes, and $\Delta\beta = \beta_i-\beta_r\neq 0$ is the difference between the propagation constants of the incident and reflected modes.

From Eq.~\eqref{eqn:cr} it follows that for uniform structures, the transition reflection $R_{\text{t}}(L)\sim |c_r|^2$ scales as $|M|^2/L^{2d} = L^{-2(d+1)}$. That is,
\begin{equation}
	R_{\text{t}} = \mathcal{O}(L^{-2(d+1)}),\quad \mbox{as}\ L\rightarrow\infty.
	\label{eqn:conv}
\end{equation}
This is confirmed by the numerical results given in the next section.

The situation is more complicated in the case of periodic structures where the phenomenon of slow light occurs near and in a band gap. In such scenarios, while approaching a flat band edge, we have that $\Delta \beta  = \beta_i - \beta_r = 2\left(\beta - \frac{\pi}{\Lambda}\right) \sim v_g $~\cite{johnson2002adiabatic,oskooi2008failure}, where $\Lambda$ is  the period of the structure and $v_g$ is the group velocity.  Also, the coupling coefficient $M$ is proportional to $ 1/{v_g}$, therefore we have that the transition reflection scales as
\begin{equation}
 	R_t \sim |c_r|^2 = \mathcal{O}(v_g^{-2(d+2)}),\quad \text{for small $v_g$},
\end{equation}
for periodic structures. ``Slow light'' corresponds to small $v_g$, hence we anticipate the need for much longer absorbers in order for the $\mathcal{O}(L^{-2(d+1)})$ decay of (\ref{eqn:conv}) to overcome this unfavorable scaling when operating close to a band edge. Such a periodic structure is the Bragg grating considered in Section~\ref{ss:slow_light}. There we observe in practice this predicted worsening in the performance of adiabatic absorbers.

\section{Numerical results}
\label{sec:num_results}
In this section, we demonstrate via numerical experiments that the asymptotic results for adiabatic absorbers discussed in the previous section are achieved in our VIE setting. We begin by considering the simple scenario of the straight uniform waveguide of Fig.~\ref{fig:strip} with an absorber of length $L$ appended to the right end. Values of $L$ from 450nm to 9450nm in increments of 450nm are considered; this equates approximately to 1 to 20 interior wavelengths. The absorption profiles are the monomials (\ref{eqn:profiles}). For this example, we observe the asymptotic behavior (\ref{eqn:conv}) of the transition reflection. Furthermore, we observe that matching the impedances by introducing magnetic conductivity improves the absorber, however only by a constant factor.

Next, we examine the behavior of adiabatic absorbers while terminating an infinitely long periodic channel, which is, in our case, the Bragg grating. First, to demonstrate the filtering behavior of the Bragg grating, we obtain the transmission spectrum for the finite grating of $N = 100$ periods. Next, we terminate the grating of $N = 50$ periods by an absorber of the same shape and with length ranging from $1\Lambda\ (\approx0.7\lambda)$ to $800 \Lambda\ (\approx 585\lambda)$, where $\Lambda = 320\rm{nm}$ is the grating's period. The simulations performed using an absorber of length $900\Lambda$ are used to generate the reference solutions. We demonstrate that near a band gap, where the group velocity goes close to zero, the effectiveness of adiabatic absorbers deteriorates. We note that this is a problem inherent to such absorbers, as well as PMLs, and is predicted by theory.

Finally, we examine the performance of absorbers of fixed length and profile to truncate a Y-branch splitter which is an oft-simulated nanophotonics structure~\cite{chrostowski2015silicon}. For all simulations performed, we use an iterative solver with tolerance $10^{-8}$ to solve the arising discrete system.

\subsection{Dielectric strip waveguide}
\label{ss:strip}
Recall the definition of the geometry for this example from Section~\ref{ss:strip_geometry}. To measure the transition reflection, we first extract the field along the central axis of the waveguide, i.e., $0\leq x\leq 13500\text{nm},\ y=250\text{nm},\ z=100\text{nm}$. The field on this axis obtained using the longest absorber ($L=9450$nm) shall be taken as the ``exact'' solution and denoted $\mathbf{E}^{\infty}$. Then the reflection coefficient, which is identified with the transition reflection up to some constant, shall be calculated as follows
\begin{equation}
	\mathtt{R} := \frac{||\mathbf{E}^{\infty}-\mathbf{E}^{(L)}||^2}{||\mathbf{E}^{\infty}||^2}.
\end{equation}
Recall we are considering the reflected power, hence the powers of 2. The norm is the $L^2$ norm, that is we have that
\begin{equation}
	||\mathbf{E}^{\infty}-\mathbf{E}^{(L)}||^2 := \int_0^X|\mathbf{E}^{\infty}(x) - \mathbf{E}^{(L)}(x)|^2\sd x.
\end{equation}

\begin{figure}[h!]
\centering
	\includegraphics[width=0.45\textwidth]{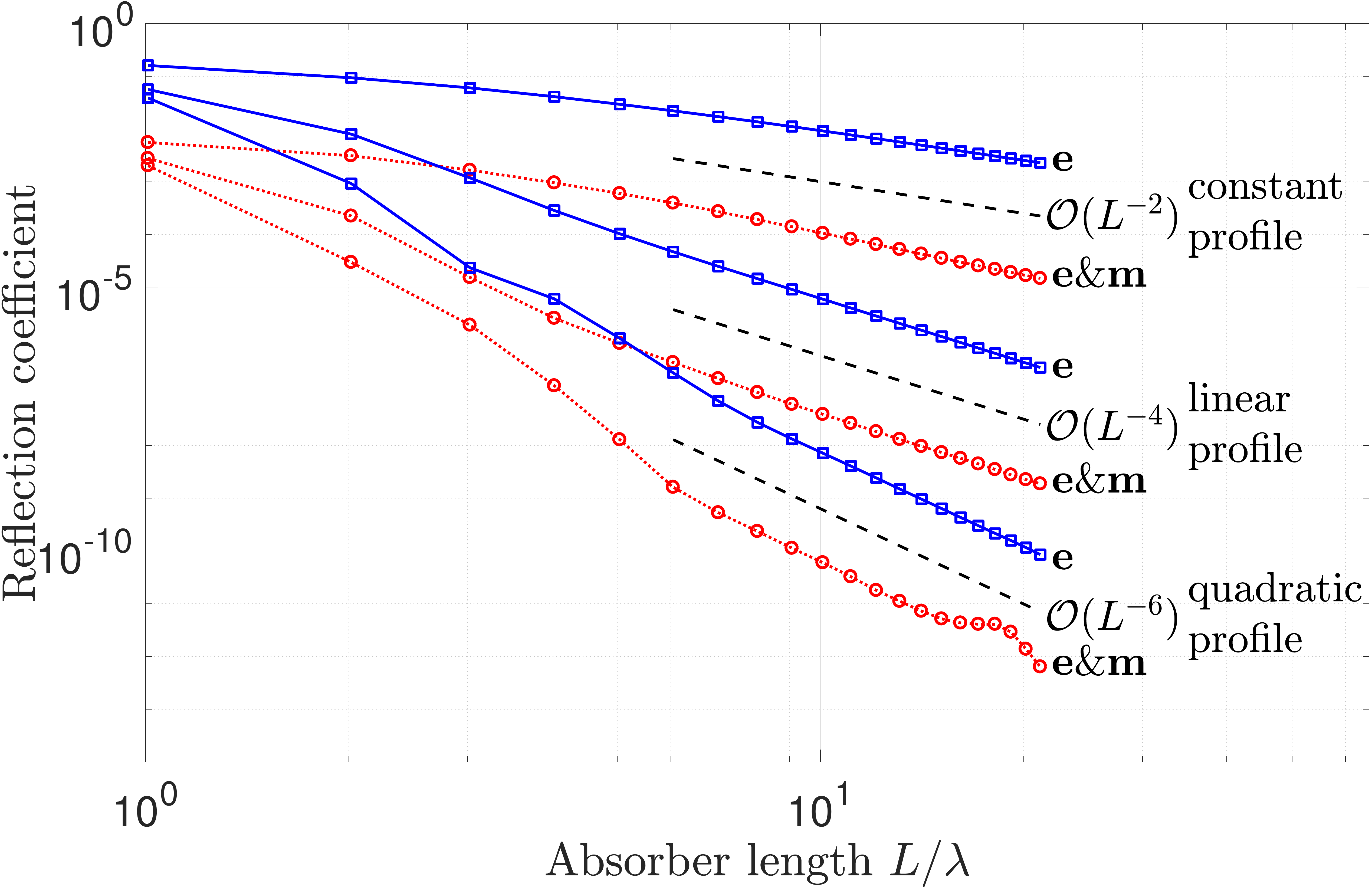}
	\caption{Reflection coefficient versus absorber length $L$ for first three monomial absorption profiles when $R_{\text{rt}}=10^{-25}$. The lines labeled $\mathbf{e}$ correspond to the case where only the electric field is solved for. The lines labeled $\mathbf{e}\&\mathbf{m}$ correspond to the case where both the electric and magnetic fields are solved for, with the impedance matched. Note that this impedance matching reduces the reflection coefficient by approximately a factor of 100. We observe that each of the first three monomials achieves the asymptotic convergence rate of $\mathcal{O}(L^{-2(d+1)})$.}
	\label{fig:1emin25}
\end{figure}

\begin{figure}[h!]
\centering
	\includegraphics[width=0.45\textwidth]{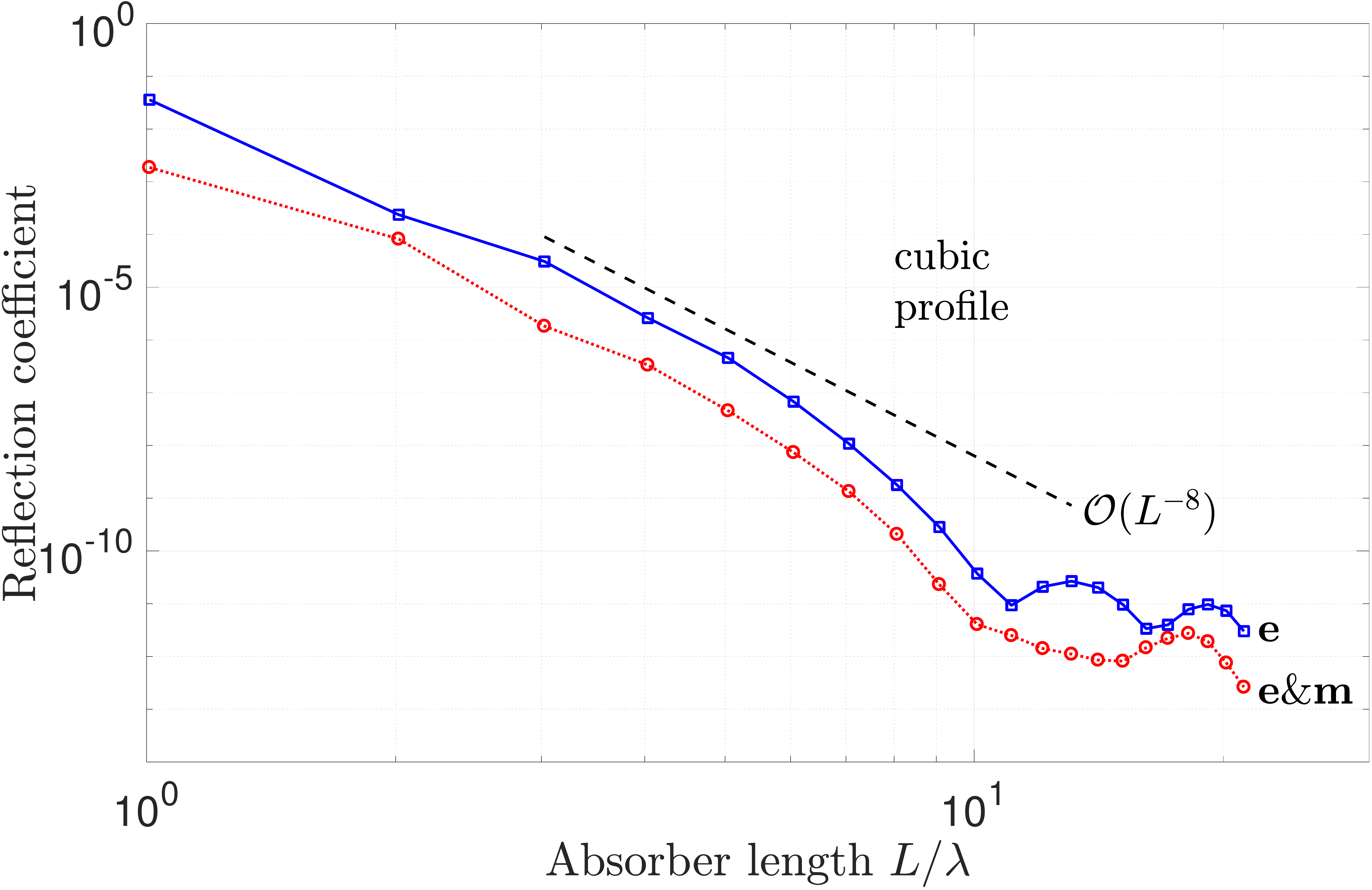}
	\caption{Reflection coefficient versus absorber length $L$ for cubic absorption profiles when $R_{\text{rt}}=10^{-25}$. The blue squares correspond to the case where only the electric field is solved for. The red circles correspond to the case where both the electric and magnetic fields are solved for, with the impedance matched. Observe that the convergence is faster than the theoretical asymptotic rate, implying the asymptotic range is not achieved for these absorber lengths. In fact, for cubic and higher order profiles, the asymptotic range is not achieved for any of the practical examples considered in this paper.}
	\label{fig:1emin25_cubic}
\end{figure}

We begin by setting the round-trip reflection to be $R_{\text{rt}}=10^{-25}$ using the relation (\ref{eqn:sigma0}). Initially we solve for the electric currents (related to $\mathbf{e}$ via (\ref{eqn:equiv_currents})) alone, that is, we do not match the impedances of the absorber and waveguide. The reflection coefficients $\mathtt{R}$ for the first three absorption profiles as functions of $L$ are shown as the lines labeled $\mathbf{e}$ (for electric) in Fig.~\ref{fig:1emin25}. 
The figure shows that the observed transition reflection agrees with the asymptotic result (\ref{eqn:conv}) for the first three monomials. The reflection for the cubic profile, as shown in Fig.~\ref{fig:1emin25_cubic}, appears not to have reached the asymptotic regime before stagnating at a value of $\mathtt{R}\approx 10^{-11}$. In this regime, it appears that the quadratic profile produces the smallest transition reflections for absorbers of length less than $8\lambda$. For absorbers of length $8\lambda$ or more, the cubic profile is superior. In general, we remark that, although increasing the polynomial degree (and hence the smoothness of the transition from waveguide to absorber) improves the asymptotic rate at which the transition reflection diminishes, it also requires ever longer absorbers in order to reach this asymptotic phase.

Next, it is interesting to observe the improvement in the absorbing layer when magnetic currents are introduced, thus allowing the impedance to be matched (albeit at the cost of doubling the number of unknowns). The lines labeled with $\mathbf{e}\&\mathbf{m}$ (for electric and magnetic currents) in Fig.~\ref{fig:1emin25} are the matched impedance counterparts of the lines labeled $\mathbf{e}$. The reflection coefficient is reduced by a factor of approximately 100, which is good, but not overwhelmingly so, and does not justify the increase in computational cost of solving for the additional magnetic currents. This reduction factor, however, is not the piece of information we are really interested in. We would like to know by how much we can reduce the absorber length to maintain the same reflection. Suppose we desire $\mathtt{R}=10^{-8}$, which is sufficiently small for practical purposes. The required absorber lengths to achieve this are shown in Table~\ref{tab:length}. Considering $R_{\text{rt}}=10^{-25}$, we see that, the reduction in required absorber length when going from unmatched impedances to matched impedances diminishes as the monomial degree increases. For $d=0$, the decrease is thousands of wavelengths, whereas when $d=3$, the decrease is little more than one wavelength. Curiously, we observe that, depending on which technique is used, either $d=2$ or $d=3$ provide the superior absorber. In any case, the small saving in the simulation domain for $d=2,3$ does not justify doubling the degrees of freedom.
\begin{table}[h!]
\centering
\begin{tabular}{|c | c | c | c | c|}
\hline
		Monomial  & \multicolumn{4}{|c|}{Absorber length $(L/\lambda)$}    \\
		\cline{2-5}
		 degree, $d$  &  \multicolumn{2}{|c|}{$R_{\text{rt}}=10^{-25}$}  & \multicolumn{2}{|c|}{$R_{\text{rt}}=10^{-10}$} \\
		 \cline{2-5}
		                       &$\mathbf{e}$  & $\mathbf{e}$\&$\mathbf{m}$ &  $\mathbf{e}$  & $\mathbf{e}$\&$\mathbf{m}$   \\
		 \hline\hline
		 0  & 13,000 & 350 & 4,500 & 230 \\
		 1  &  49 & 14 & 32 & 9.0 \\
		 2  & 9.5 & 5.1 & 7.1 & 4.8 \\
		 3 & 7.2 & 5.9 & 6.2 & 6.1 \\
		 \hline
	
\end{tabular}
\caption{Absorber length (in units of number of wavelengths) required to obtain $\mathtt{R} =10^{-8}$. We choose such a value for $\mathtt{R}$ since it is sufficiently small for practical purposes. Some of the values for $d=0,1$ have been extrapolated from Fig.~\ref{fig:1emin25} and Fig.~\ref{fig:1emin10}.}
\label{tab:length}
\end{table}

\begin{figure}[h!]
\centering
	\includegraphics[width=0.45\textwidth]{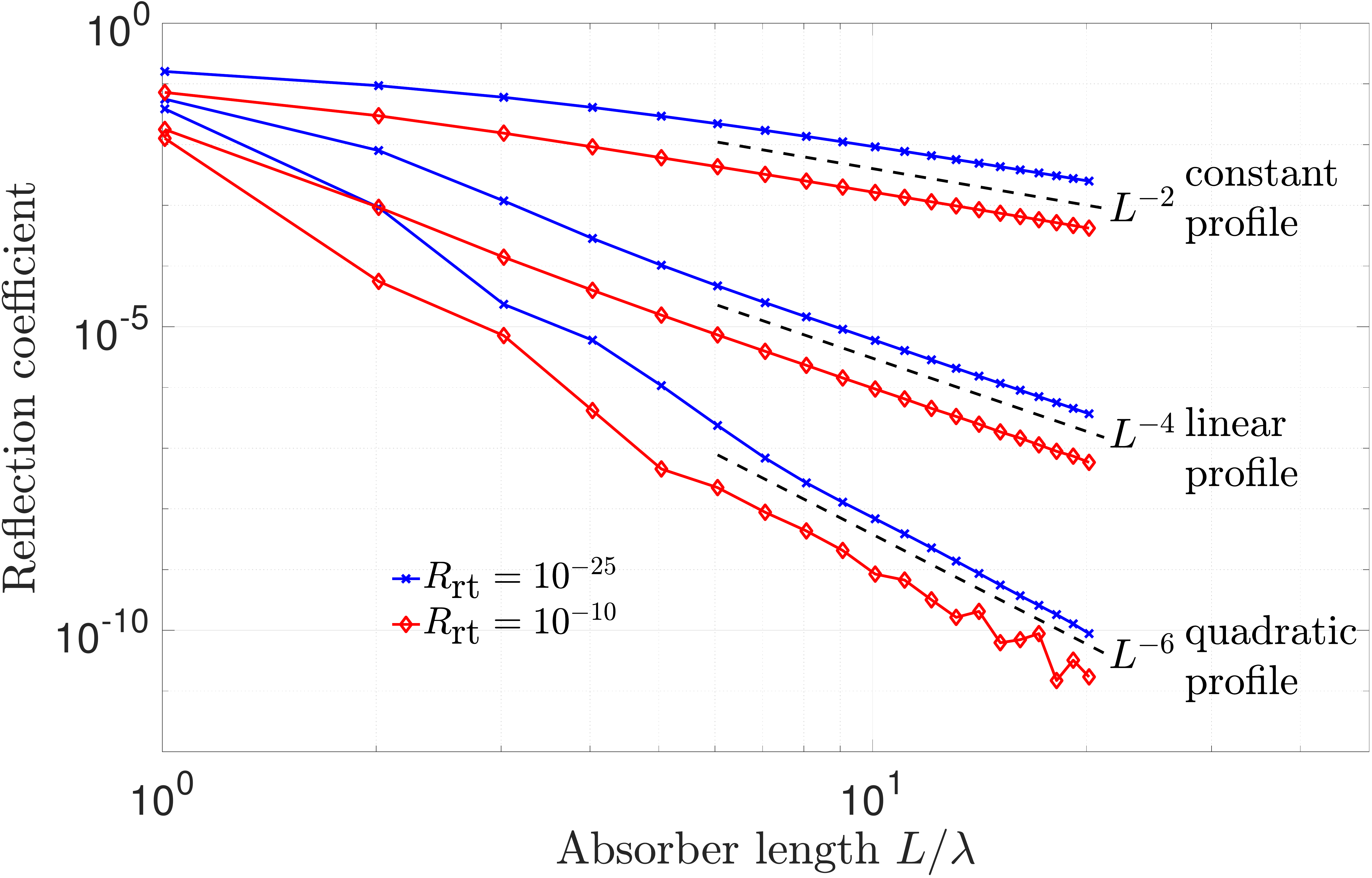}
	\caption{Reflection coefficient versus absorber length $L$ for first three monomial absorption profiles for two different round-trip reflections: $R_{\text{rt}}=10^{-25}$ (blue crosses) and $R_{\text{rt}}=10^{-10}$ (red diamonds). An order of magnitude reduction in the reflection coefficient is achieved by decreasing the imposed round-trip reflection in this way.}
	\label{fig:1emin10}
\end{figure}

A cheaper way to reduce the transition reflection, and hence the required size of the absorber, is to reduce the imposed round-trip reflection $R_{\text{rt}}$. Consider $R_{\text{rt}}=10^{-10}$: the corresponding results are shown in Fig.~\ref{fig:1emin10} and Table~\ref{tab:length}. To obtain $\mathtt{R}=10^{-8}$ with a quadratic profile, say, we require an absorber of length 7.1$\lambda$ for $R_{\text{rt}}=10^{-10}(\mathbf{e}$), compared to 9.5$\lambda$ for $R_{\text{rt}}=10^{-25}(\mathbf{e}$), and 5.1$\lambda$ for $R_{\text{rt}}=10^{-25}(\mathbf{e}$\&$\mathbf{m}$). The reduction is not quite as large as observed when introducing magnetic currents (two wavelength less in fact), but it is significant and we do not have to double the number of unknowns. 

It is also worth noting that in all our convergence graphs, the reflection coefficient stagnates at some small value. This is due to the difference between the phases of the round-trip reflections contained in $\mathbf{E}^{(L)}$ and $\mathbf{E}^{\infty}$, for each $L$. Therefore, we would anticipate that this small stagnation value should be close to the enforced round-trip reflection. Indeed, this is the case when $R_{\text{rt}}=10^{-10}$ in Fig.~\ref{fig:1emin10}. However, this is not the case when $R_{\text{rt}}=10^{-25}$ in Fig.~\ref{fig:1emin25}. In the latter scenario, this is because the numerical discretization error and iterative solver tolerance are greater than the enforced round-trip reflection.

\begin{figure}[h!]
\centering
	\includegraphics[width=0.45\textwidth]{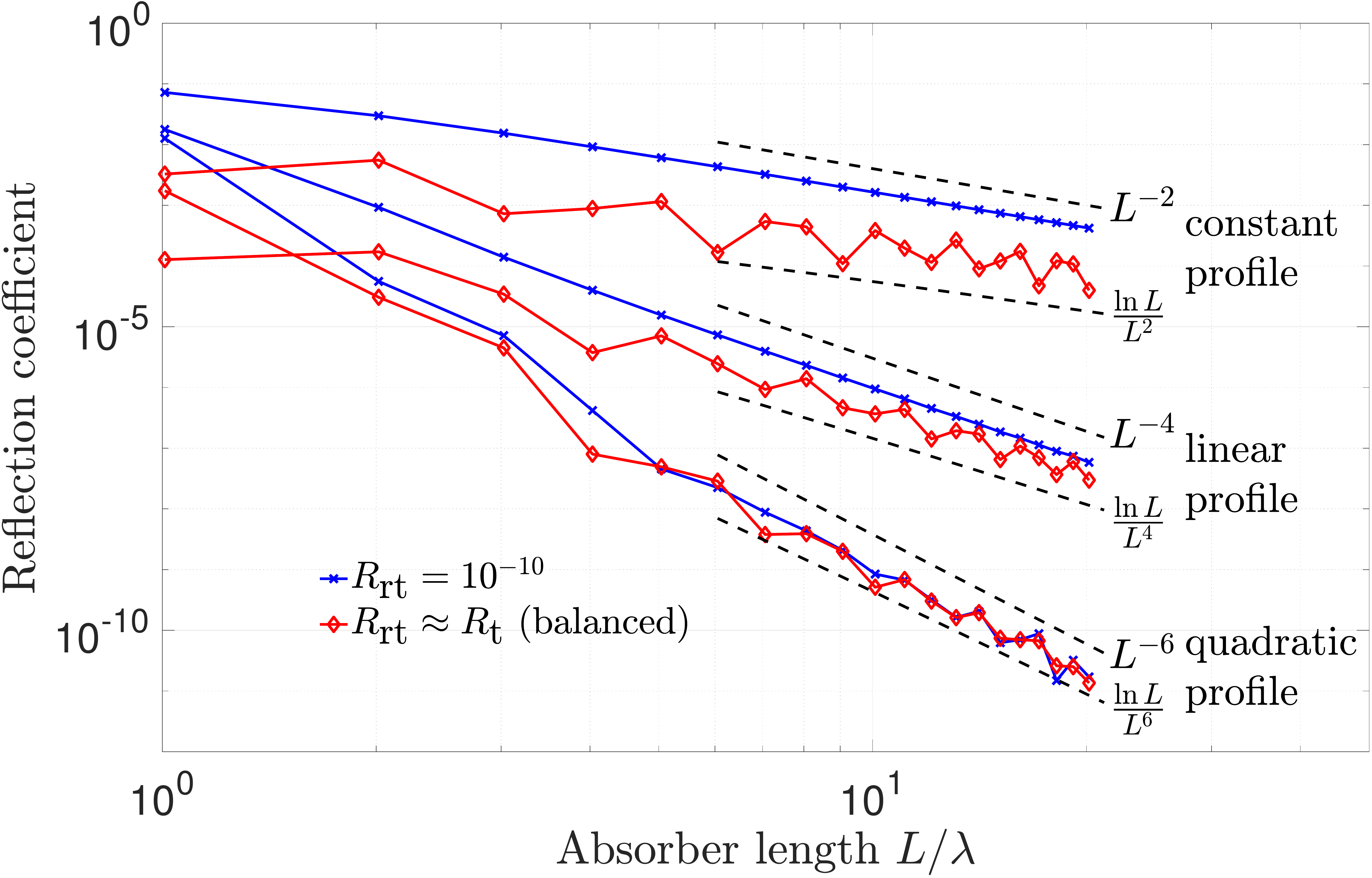}
	\caption{Reflection coefficient versus absorber length $L$ for first three monomial profiles for two different round-trip reflections: $R_{\text{rt}}=10^{-10}$ (blue crosses) and $R_{\text{rt}}=C_{\text{opt}}L^{-2(d+1)}$ (red diamonds). The second case is equivalent to balancing the round-trip reflection with the transition reflection. Observe the improvement achieved; this gain diminishes as the polynomial degree increases.}
	\label{fig:balanced_all}
\end{figure}

The next step to optimize the adiabatic absorber is to balance the round-trip and transition reflections for a given length $L$. This entails enforcing the round-trip reflection to also follow the power law (\ref{eqn:conv}). This leads to larger values of $R_{\text{t}}$ for smaller $L$, hence a smaller $\sigma_0 \propto -\ln(R_{\text{rt}})$, and thus a smaller transition reflection. Fig.~\ref{fig:balanced_all} shows the result of balancing $R_{\text{rt}}$ and $R_{\text{t}}$ for a constant absorption profile. We observe that the reduction in the transition reflection is substantial for small $L$. For larger $L$ the reduction is less dramatic; this is due to an additional factor of $\ln R_{\text{rt}}\sim \ln L$ that now appears in the asymptotic convergence rate.
Note further that the line is jagged rather than straight. This is due to the interference of the now similar size round-trip and transition reflections. In Fig.~\ref{fig:balanced_all} are shown the reflection coefficients for the first three monomial profiles. The improvement gained by balancing $R_{\text{rt}}$ and $R_{\text{t}}$ appears to reduce as the polynomial degree of the absorption profile is increased. Finally, we note that, although balancing these two reflections gives a reduction in transition reflection, it requires some trial and error in order to choose the optimal constant $C_{\text{opt}}$ in the imposed power law for the round-trip reflection $R_{\text{t}}=C_{\text{opt}}L^{-2(d+1)}$. For practical purposes, one would wish to perform such an optimization over all problem parameters such as wavelength and refractive index. This is a non-trivial task and, as can be seen, yields little gain. Further exploration of such an optimzation is left to future work.

\subsection{Bragg grating}
\label{ss:slow_light}
\begin{figure}[h!]
\centering
		\includegraphics[width=0.45\textwidth]{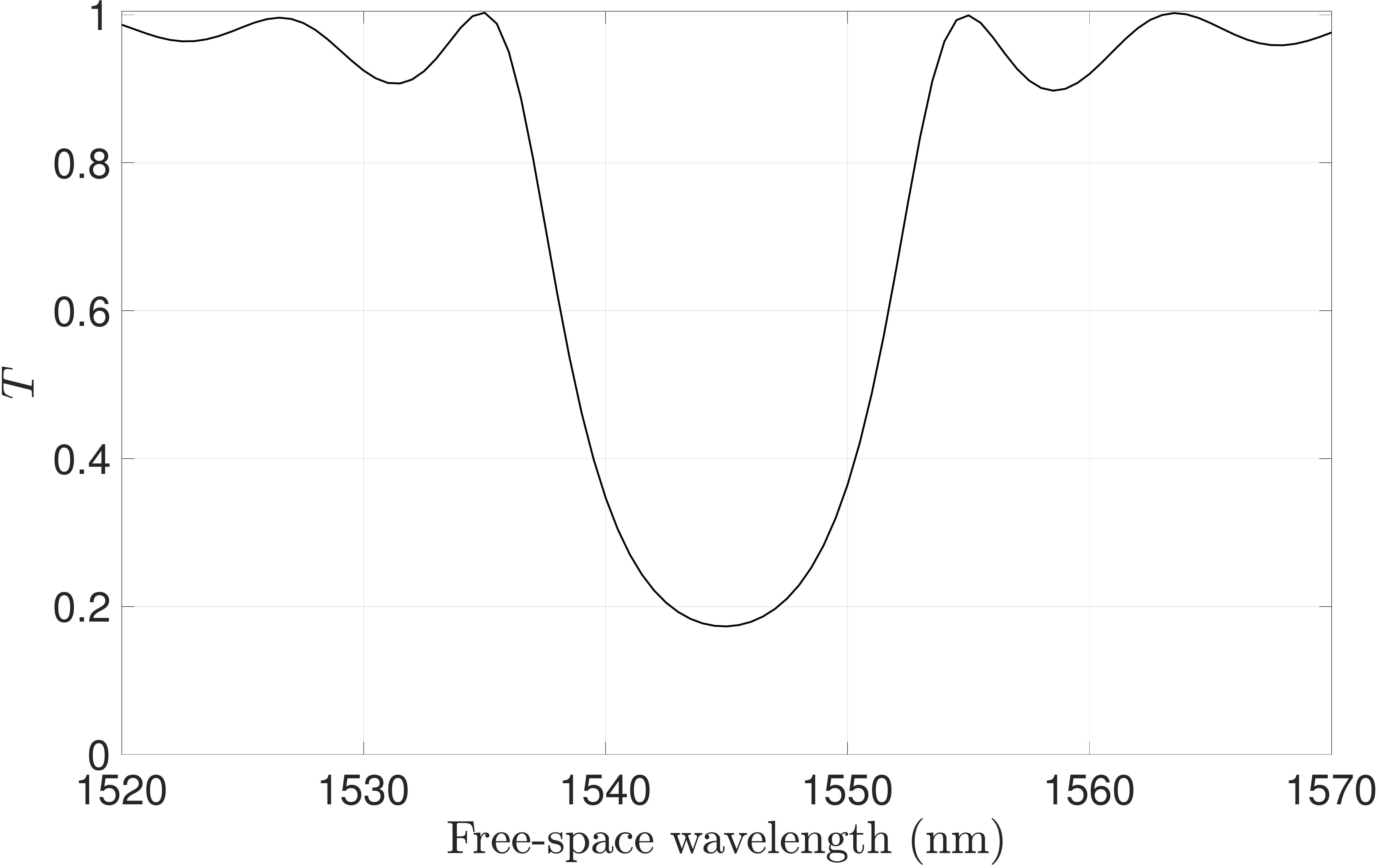}
	\caption{Transmission through the Bragg grating of length 100 periods. The Bragg wavelength is 1545nm and the band gap is approximately 20nm wide.}
	\label{fig:Bragg_trans}
\end{figure}

In this section, we consider the Bragg grating structure which generates the phenomenon of \textit{slow-light} for certain wavelengths; the wavelengths for which this occurs are effectively filtered out of an input signal. We begin by demonstrating the filtering behavior of a Bragg grating of finite length, as described in Section~\ref{ss:Bragg_geometry}. In order to do so, we perform simulations on the setup in Fig.~\ref{fig:Bragg} over the free-space wavelength range [1520,1570]nm at a sampling resolution of 0.5nm. A quadratic absorber of length 2.2$\mu$m ($\approx 5\lambda$) is used on either end. For each wavelength, the transmission $T$ through the Bragg grating is defined as the integrated square of the electric field over a voxel-wide $(y,z)$-slice of the structure:
\begin{equation}
	T = \iint |\mathbf{E}|^2\sd y \sd z.
\end{equation}
This chosen slice must be located after the Bragg grating terminates and before the absorber begins. The normalized transmission is plotted in Fig.~\ref{fig:Bragg_trans}. We observe a region in which the transmission drops significantly; this is known as the \textit{band gap}, which here is approximately 20nm wide with its center at a \textit{Bragg wavelength} of 1545nm. At the wavelengths corresponding to the band gap, the reflections from the modulations of the Bragg interfere constructively and hence lead to the light being strongly reflected. In this way, a Bragg grating acts to filter out these wavelengths from an input signal. The drop in transmission is related to the number of periods in the grating, and for the very long gratings used in practice, the transmission in the band gap is much closer to zero than for the short example considered here. 

If one were to analyze the group velocity, $v_g$, it would be seen to be positive away from the band gap, approach zero at the band gap edge, then be negative within the band gap (see \cite{joannopoulos2011photonic} for in-depth details on band gaps in periodic structures). Now suppose we were to introduce an adiabatic absorber within the periodically modulated region of the Bragg, as depicted in the inset in Fig.~\ref{fig:Bragg_convergence_1520}. As discussed at the end \S\ref{sec:adiabatic}, the transition reflection from this absorber would be dominated by the $\mathcal{O}(v_g^{-2(d+2)})$ term as we go past a band gap edge. Thus we would require extremely long absorbers before returning to our asymptotic (in $L$) convergence rate of $\mathcal{O}(L^{-2(d+1)})$. 

From a purely physical point of view, we should expect such a deterioration in performance of absorbing layers in this scenario. The aim of employing an absorbing layer is to allow the truncation of the domain \textit{without incurring reflections}. However, the slow light phenomenon described above occurs precisely \textit{due to the reflections} from all the way along the Bragg grating. By terminating the structure with an absorber, we lose these important reflections and, more importantly, we lose periodicity which is essential for the propagation of Bloch waves. Therefore, we anticipate that extremely long absorbers will be required to retain a sufficient number of these reflections in order to mimic the field within the infinite periodic structure. 

We proceed by performing such a set of simulations in order to model the infinite Bragg grating and thereby observe the aforementioned behavior of the absorber. That is, we terminate the periodically varying region with an absorber of the same shape, as shown in Fig.~\ref{fig:Bragg_convergence_1520}. The absorption profiles are again the monomials~\eqref{eqn:profiles}.
\def\nn{13}
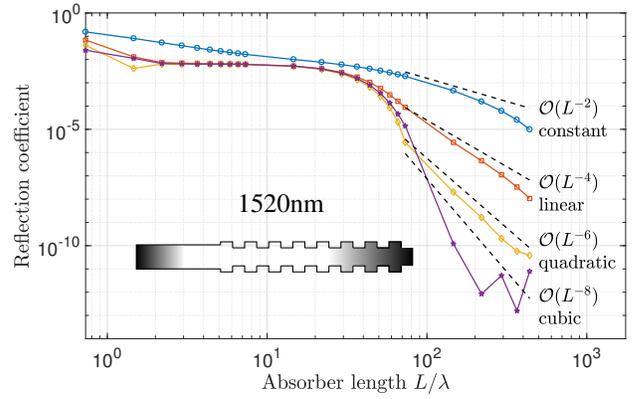
\begin{figure}
  \centering   
  \begin{overpic}[width=0.45\textwidth]{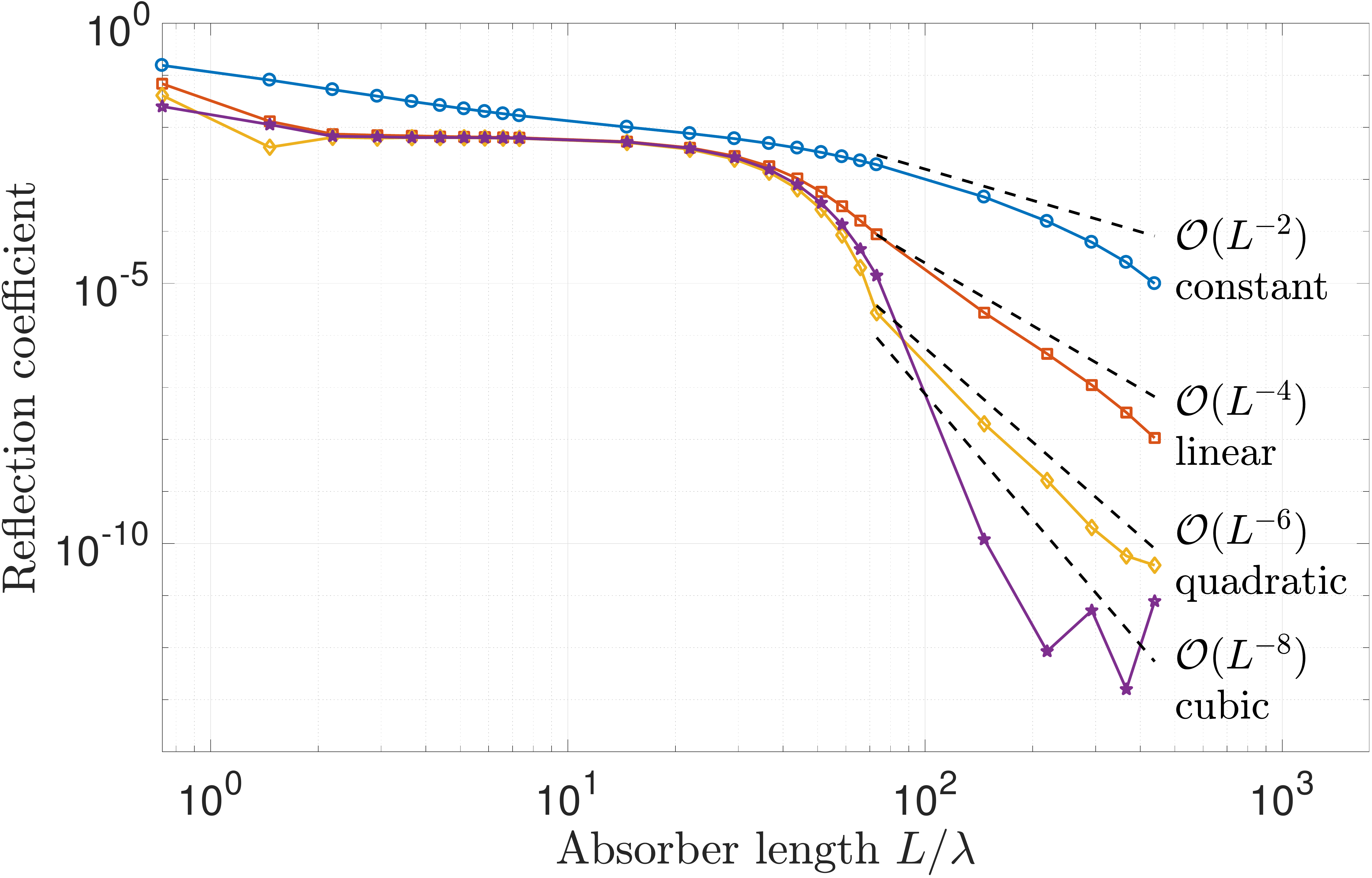}
     \put(20,20){\begin{tikzpicture}[scale=0.16]
	
	\shade[left color=black,right color=white] (-6,2) rectangle (-2,0);
	
	\shade[left color=white,right color=black] (9+1,1.75)--(9+2,1.75)--(9+2,2.25)--(9+3,2.25)--(9+3,1.75)--(11+1,1.75)--(11+2,1.75)--(11+2,2.25)--(11+3,2.25)--(11+3,1.75)--(13+1,1.75)--(13+2,1.75)--(13+2,2.25)--(13+3,2.25)--(16,1.75)--(17,1.75)--(17,0.25)--(16,0.25)--(13+3,-0.25)--(13+2,-0.25)--(13+2,0.25)--(13+1,0.25)--(13+1,-0.25)--(13,-0.25)--(13,0.25)--(12,0.25)--(12,-0.25)--(11,-0.25)--(11,0.25)--(10,0.25)--(10,-0.25)--cycle;
	
	\draw (1,2) -- (1,2.25) -- (2,2.25) -- (2,1.75);
	\draw (1,0) -- (1,-0.25)-- (2,-0.25)-- (2,0.25);
	
	\foreach \i in {1,3,5,7,9,11,13,\nn}
	{
		\draw (\i+1,1.75)--(\i+2,1.75)--(\i+2,2.25)--(\i+3,2.25)--(\i+3,1.75);
		\draw (\i+1,0.25)--(\i+2,0.25)--(\i+2,-0.25)--(\i+3,-0.25)--(\i+3,0.25);
	}
	
	

	\draw (1,0) -- (-6,0) -- (-6,2) -- (1,2);
	
	\draw (6,7) node[anchor=north]{1520nm};
	
	\end{tikzpicture}}  
  \end{overpic}
\caption{Reflection coefficient versus absorber length $L$ for monomial profiles with the round-trip reflection set to $R_{rt} = 10^{-10}$. The Bragg grating is excited at 1520nm free-space wavelength. The asymptotic convergence rates are eventually achieved for the first three monomials.}
	\label{fig:Bragg_convergence_1520}
\end{figure}

First, we excite the system at a free-space wavelength of 1520nm, away from the band gap edge, and thus the group velocity is relatively large and positive. The round-trip reflection is fixed at $R_{rt} = 10^{-10}.$ The reflection coefficient $\mathrm{R}$ for the four monomial profiles is shown in Fig.~\ref{fig:Bragg_convergence_1520}. Immediately apparent is a clear stagnation in the transition reflection for absorbers up to approximately 40$\lambda$. Beyond this point, the transition reflections converge towards zero and we achieve close to the asymptotic convergence rate $\mathcal{O}(L^{-2(d+1)})$ for constant, linear, and quadratic profiles. 

\begin{figure}
  \centering   
  \begin{overpic}[width=0.45\textwidth]{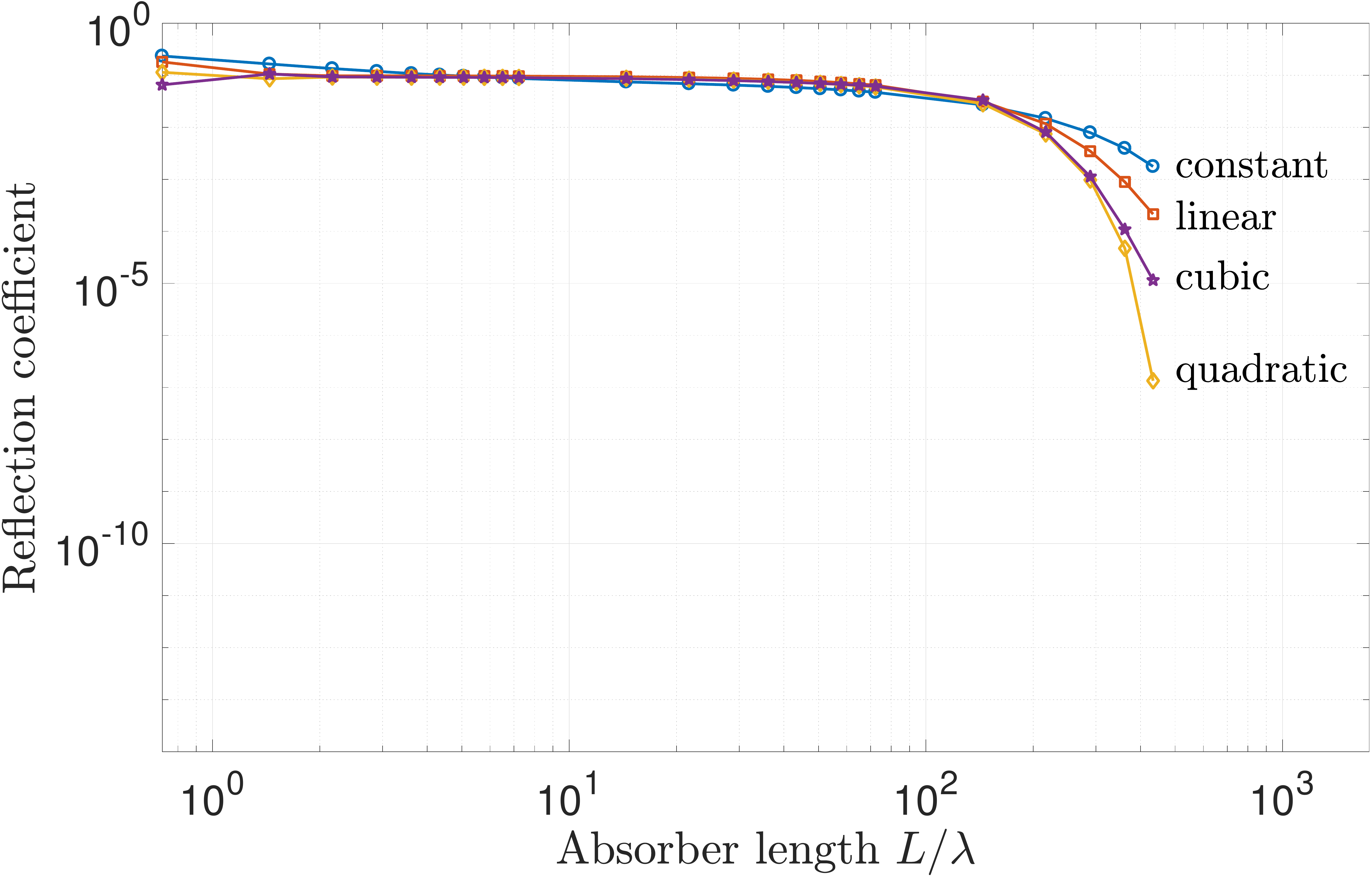}
     \put(20,20){\begin{tikzpicture}[scale=0.16]
	
	\shade[left color=black,right color=white] (-6,2) rectangle (-2,0);
	
	\shade[left color=white,right color=black] (9+1,1.75)--(9+2,1.75)--(9+2,2.25)--(9+3,2.25)--(9+3,1.75)--(11+1,1.75)--(11+2,1.75)--(11+2,2.25)--(11+3,2.25)--(11+3,1.75)--(13+1,1.75)--(13+2,1.75)--(13+2,2.25)--(13+3,2.25)--(16,1.75)--(17,1.75)--(17,0.25)--(16,0.25)--(13+3,-0.25)--(13+2,-0.25)--(13+2,0.25)--(13+1,0.25)--(13+1,-0.25)--(13,-0.25)--(13,0.25)--(12,0.25)--(12,-0.25)--(11,-0.25)--(11,0.25)--(10,0.25)--(10,-0.25)--cycle;
	
	\draw (1,2) -- (1,2.25) -- (2,2.25) -- (2,1.75);
	\draw (1,0) -- (1,-0.25)-- (2,-0.25)-- (2,0.25);
	
	\foreach \i in {1,3,5,7,9,11,13,\nn}
	{
		\draw (\i+1,1.75)--(\i+2,1.75)--(\i+2,2.25)--(\i+3,2.25)--(\i+3,1.75);
		\draw (\i+1,0.25)--(\i+2,0.25)--(\i+2,-0.25)--(\i+3,-0.25)--(\i+3,0.25);
	}
	
	

	\draw (1,0) -- (-6,0) -- (-6,2) -- (1,2);
	
	\draw (6,7) node[anchor=north]{1538nm};
	
	\end{tikzpicture}}  
  \end{overpic}
\caption{Reflection coefficient versus absorber length $L$ for monomial profiles with the round-trip reflection set to $R_{rt} = 10^{-10}$. The Bragg grating is excited at 1538nm free-space wavelength.}
	\label{fig:Bragg_convergence_1538}
\end{figure}
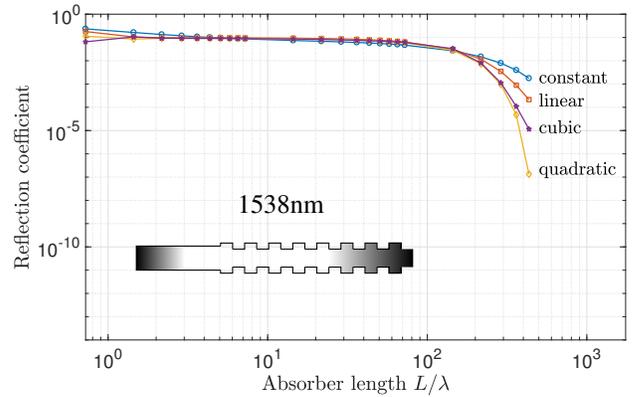

Next, we excite the grating close to the band gap edge, at a free-space wavelength 1538nm, corresponding to a lower group velocity. Comparing Fig.~\ref{fig:Bragg_convergence_1538} with Fig.~\ref{fig:Bragg_convergence_1520}, we can clearly see that reflections worsen dramatically when the system is excited near the band gap edge. Moreover, for all the profiles, the asymptotic regime is not reached for the examined range of absorber lengths, and the higher-order profiles become superior only for very long absorbers with length greater than approximately 200 wavelengths. In fact, note that the quadratic absorber is superior to the cubic absorber even for the longest absorber considered. At this wavelength, it appears that, by extrapolating from the results, a quadratic absorber of length approximately 900-1000 wavelengths would be required to provide adequately small transition reflections, and hence well-approximate the infinite Bragg grating.

\begin{figure}[ht!]
	\centering
	\includegraphics[width=0.45\textwidth]{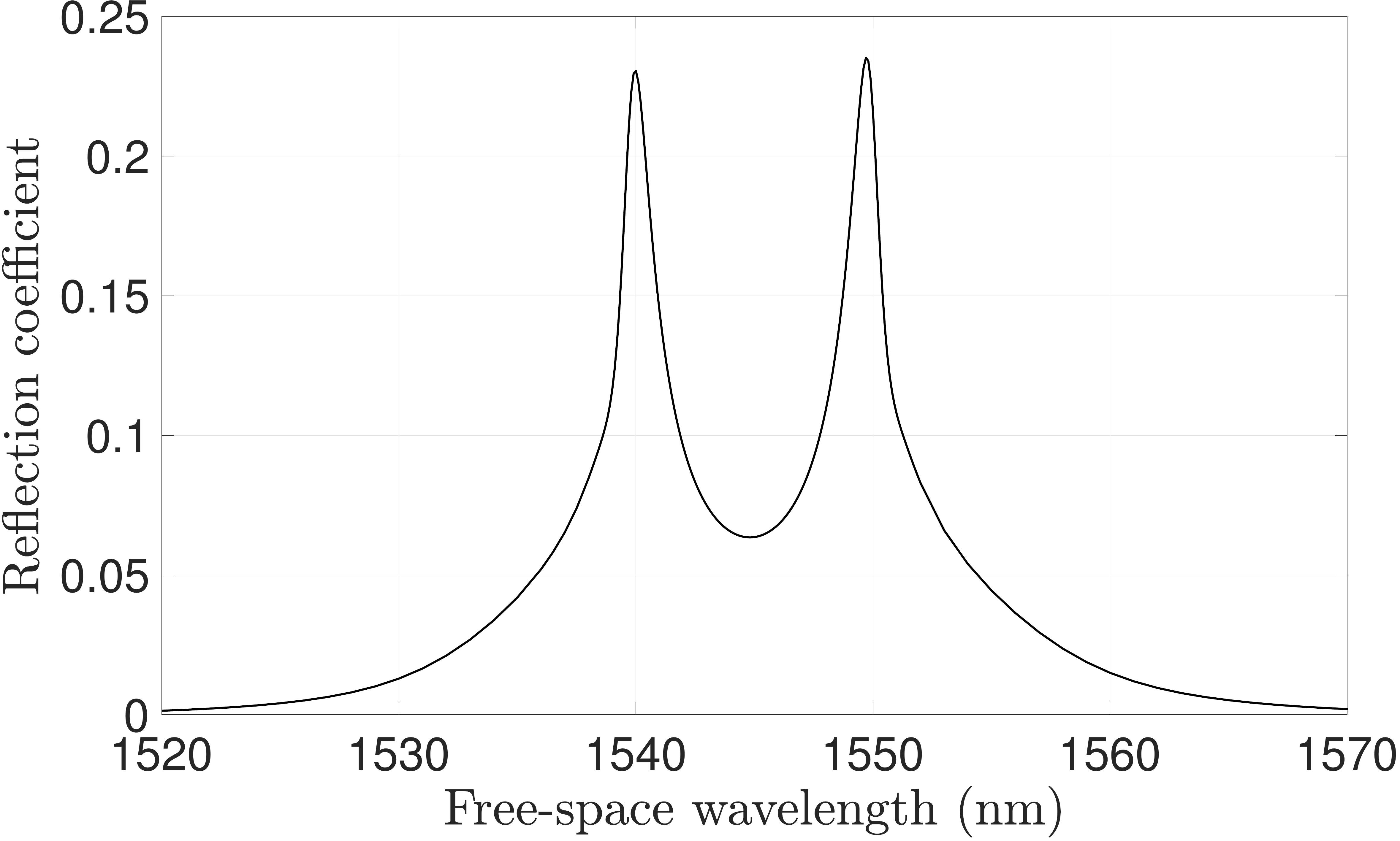}
	\caption{The reflection coefficient versus free-space wavelength for an absorber inside the ``infinite'' Bragg grating. The adiabatic absorber has a quadratic profile and length 50 periods ($\approx 36\lambda$). Comparing to Fig.~\ref{fig:Bragg_trans}, observe that the reflection from the absorber is large in and close to the band gap, with the peaks corresponding to the band gap edges where the group velocity changes sign.}
	\label{fig:R_w_quad}
\end{figure}

We conclude this section on the Bragg grating by exploring the dependency of this growth in reflection coefficient as a function of wavelength. We fix the absorber length at 50 periods ($\approx 36\lambda$) and consider the quadratic profile, and calculate the reflection coefficient at each free-space wavelength in the range [1520,1570]nm, with the reference solutions being calculated with quadratic absorbers of length 650 periods. The results are shown in Fig.~\ref{fig:R_w_quad}. We observe that the reflection has two maxima, at 1540nm and 1550nm which correspond to the band gap edges which can be observed in Fig.~\ref{fig:Bragg_trans}. These are the points where the group velocity $v_g$ passes through zero as it changes sign. Recall that the CMT predicts that this curve has the shape $\mathcal{O}(v_g^{-8})$ (from substituting $d=2$ into $\mathcal{O}(v_g^{-2(d+2)})$). In light of these large reflections near or across the band gap, there now appears to be a large amount of room for balancing predicted reflections of size  $\mathcal{O}(v_g^{-2(d+2)})$ with the round-trip reflections in order to optimize the absorber (in a similar way to the balancing of $R_{\text{t}}$ and $R_{\text{rt}}$ for the strip waveguide in Section~\ref{ss:strip}). We expect that such a balancing could lead to a significant reduction in the transition reflection, however the optimized absorbers would still have to be much longer than those for wavelengths away from the band gap due to the fact that reflections from far down the Bragg grating are important in approximating the infinite Bragg when we are near the band gap.

For Bragg grating applications, typically the entire finite grating is simulated, as was done at the beginning of this section. Therefore, this difficult behavior of absorbing layers in periodic media can be avoided. However, there are scenarios, such as photonic crystals~\cite{joannopoulos2011photonic}, where simulating the propagation of light through infinite periodic structures is of interest. For such cases, it is interesting to further understand the behavior of these absorbers in the slow-light regime in order to optimize their performance. Such a study and optimization shall be presented separately since it is not directly pertinent to the majority of nanophotonics applications.

\subsection{Y-branch splitter}
\label{ss:Y_branch}
\begin{figure}[ht!]
	\centering
	\includegraphics[width=0.5\textwidth,trim={11cm 12cm 5cm 11cm},clip]{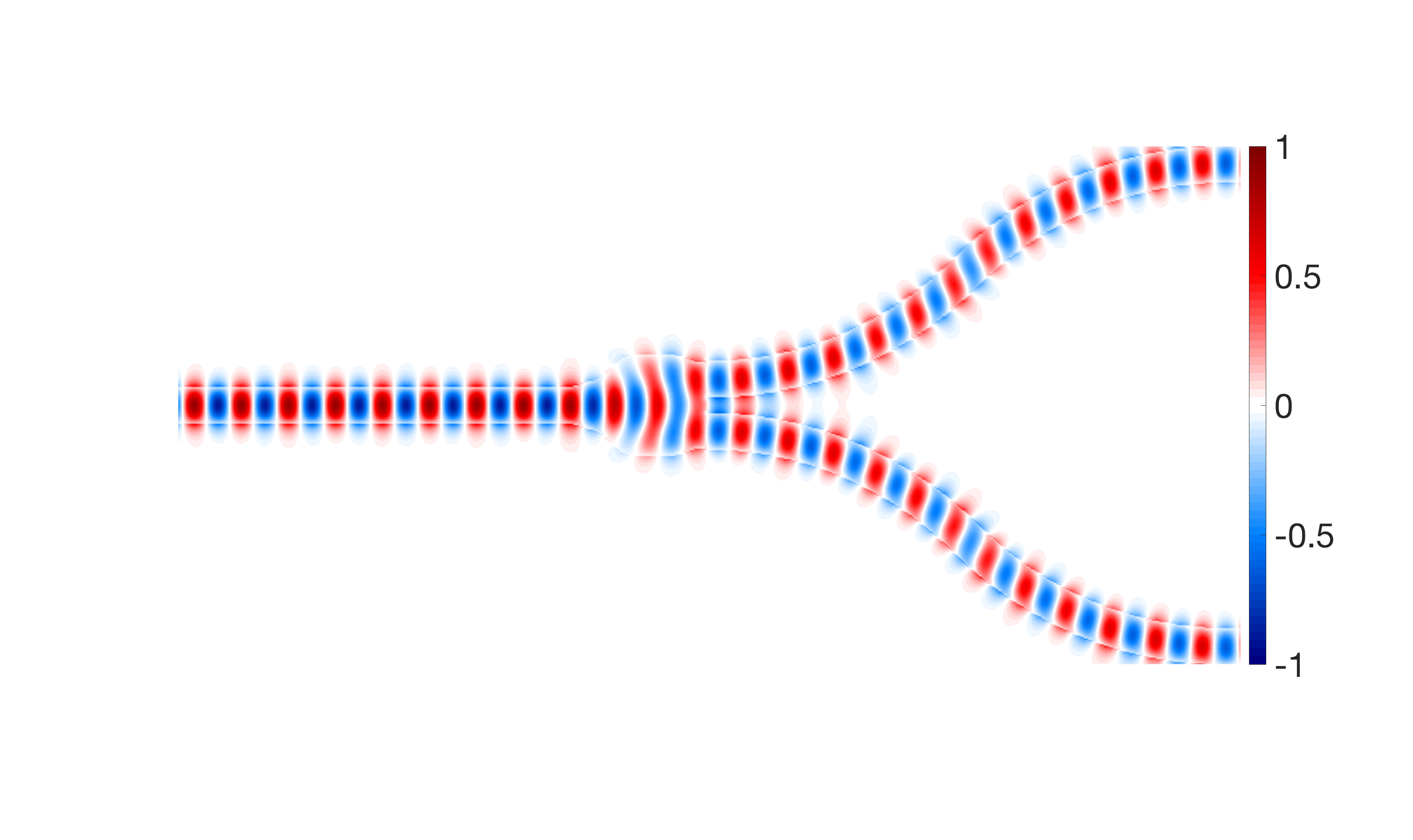}
	\caption{Real part of $\mathbf{E}_y$ (in-plane) for a silicon Y-branch splitter with SiO$_2$ cladding at 1550nm.}
	\label{fig:Y_branch_field}
\end{figure}

As the final example, we consider a practical nanophotonics simulation: the propagation of a guided mode through a Y-branch splitter, depicted in Fig.~\ref{fig:Y_branch_field}. Simulations are useful tools for optimizing the design of such structures. Indeed, the particular geometry used here is taken from \cite{zhang2013compact} where numerical simulations are used to create this low-loss design.

In our simulation, the structure is excited at the left end by a $y$-polarized Gaussian beam, establishing a guided mode in the straight waveguide which is then split at the Y-branch junction. To perform this simulation, each of the three branches is truncated with an adiabatic absorber. We choose these absorbers to have quadratic profiles and to be of length 2.2$\mu$m ($\approx 5\lambda$). We saw for the strip waveguide that this absorber with $R_{\text{rt}}=10^{-10}$ yielded a reflection coefficient of approximately $5\times10^{-8}$ (see Fig.~\ref{fig:1emin10}). 
\label{ss:y_branch}

\begin{figure}[ht!]
	\centering
	\includegraphics[width=0.5\textwidth,trim={11cm 12cm 5cm 11cm},clip]{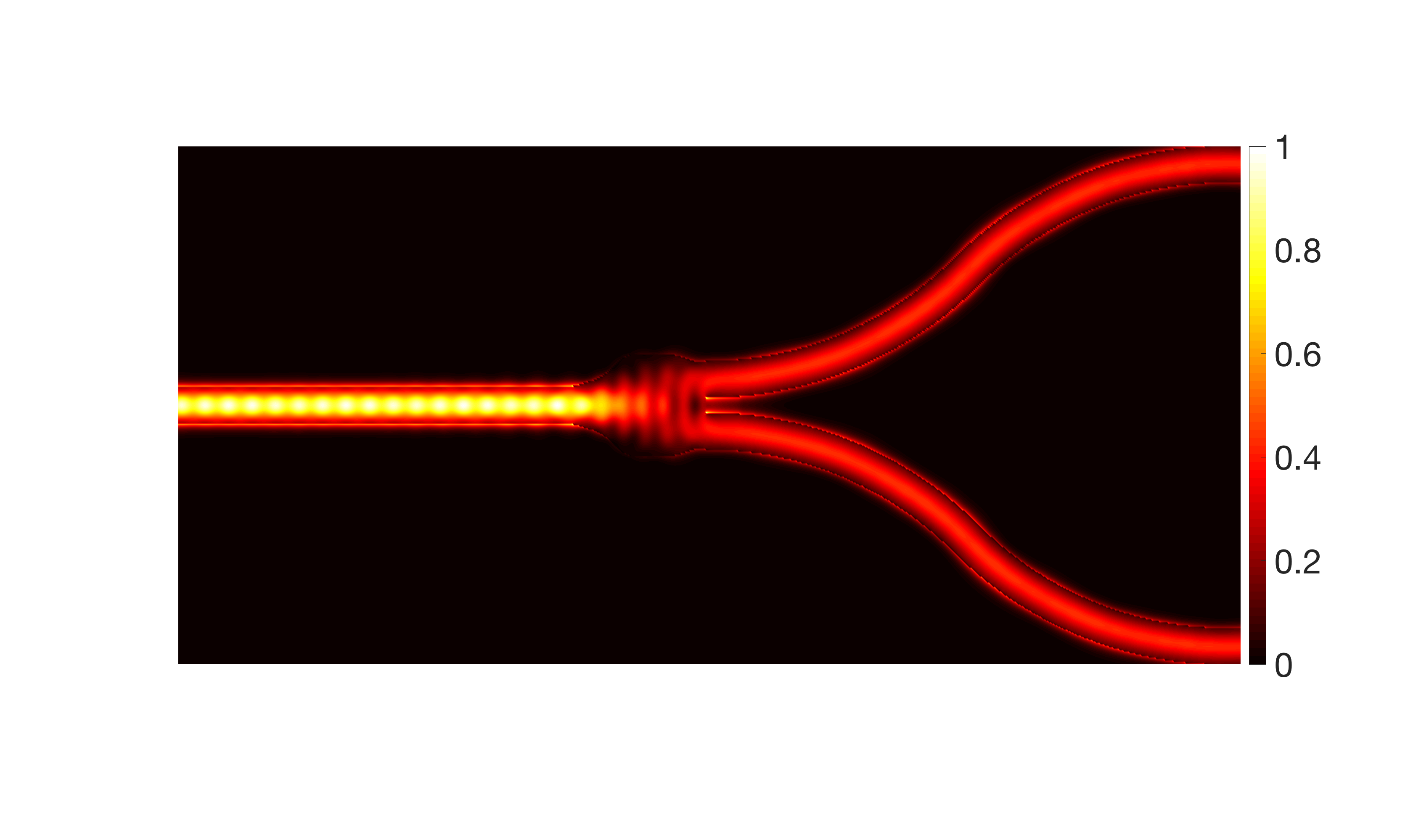}
	\caption{$|\mathbf{E}|^2$ for a silicon Y-branch splitter with SiO$_2$ cladding at 1550nm.}
	\label{fig:Y_branch_power}
\end{figure}

The field produced by the simulation is shown in figures \ref{fig:Y_branch_field} and \ref{fig:Y_branch_power}. In Fig.~\ref{fig:Y_branch_field}, we observe the real part of the in-plane field. The highly confined guided mode propagating from the left is clearly evident. As is passes through the junction, it is split into two guided modes propagating along the curved branches with very little energy being scattered outside of the structure. 
In Fig.~\ref{fig:Y_branch_power} the square of the field's magnitude is shown. Here one can see that fairly substantial reflections from the junction are propagating back down the left waveguide. This suggests that there is still some room for improvement when it comes to optimizing this Y-branch geometry. Ideally, precisely half of the energy of the incident mode would propagate down each of the two curved branches, thus requiring no reflection or scattering from the junction.

\begin{figure}[ht!]
	\centering
	\includegraphics[width=0.5\textwidth,trim={11cm 12cm 5cm 8cm},clip]{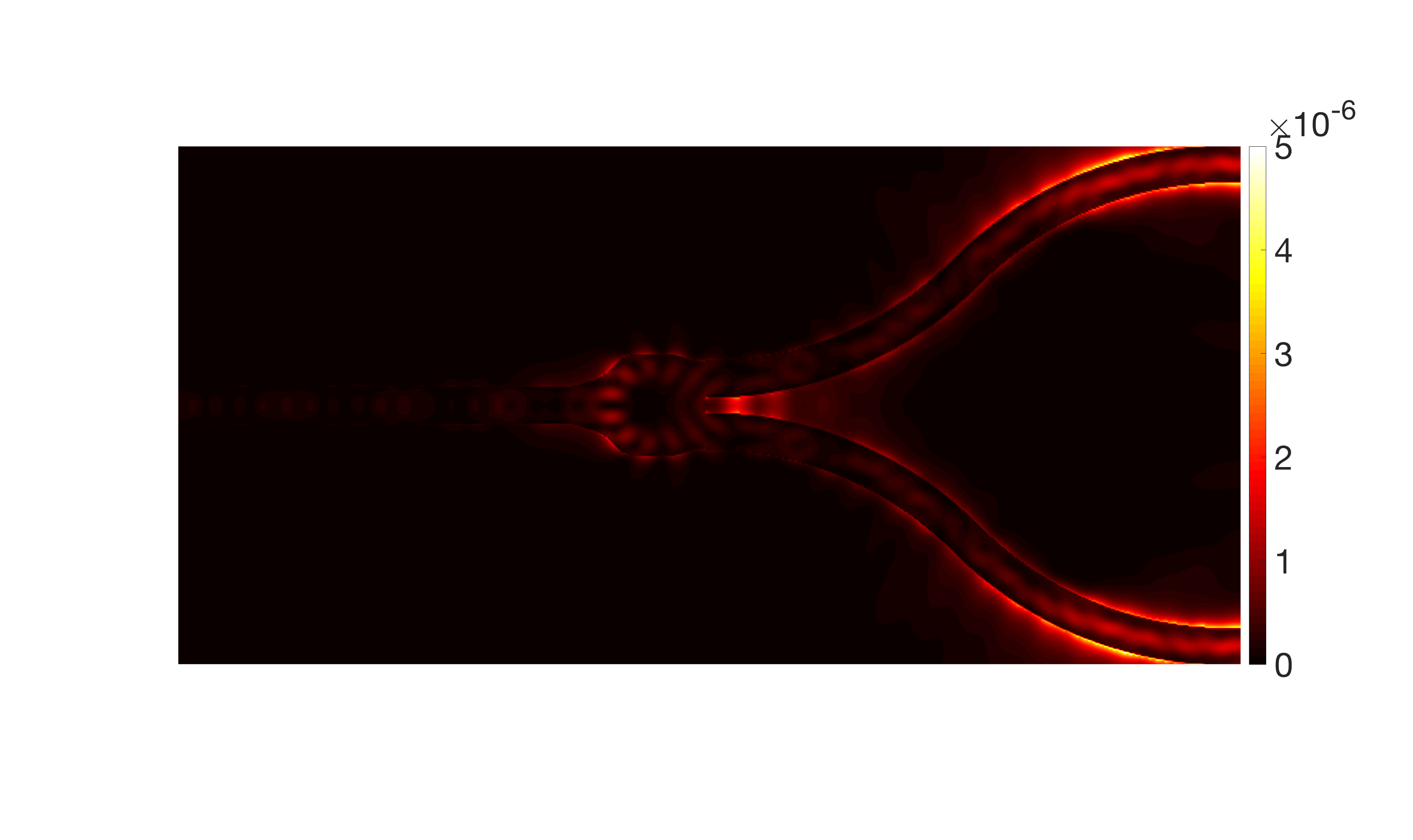}
	\caption{Relative difference in solution ($|\mathbf{E}|^2$) obtained using absorbers of length 5$\lambda$ and $20\lambda$. This difference can be attributed to the reflection from the $5\lambda$-length absorbers.}
	\label{fig:Y_branch_error}
\end{figure}

It is interesting to look at how effective the adiabatic absorbers are in this simulation. We do this by running the same simulation but now with absorbers of length 8.8$\mu$m ($\approx 20\lambda$). The field produced by this simulation, $\mathbf{E}_{20\lambda}$, is used as the reference solution to which we compare the field from the 2.2$\mu$m absorber simulation, $\mathbf{E}_{5\lambda}$. The relative difference 
\[
	\frac{|\mathbf{E}_{20\lambda}-\mathbf{E}_{5\lambda}|^2}{\max(|\mathbf{E}_{20\lambda}|)^2}
\]
is shown in Fig.\ref{fig:Y_branch_error}. 
This difference is similar to that analyzed previously and can be attributed to the reflection from the $2.2\mu$m absorbers. We observe the that the error is largest at the right ends of the curved branches and reaches a maximum of $5\times 10^{-6}$ which is substantially larger than the $5\times 10^{-8}$ observed for the straight strip waveguide of Section\ref{ss:strip}. This increase is due to the oblique propagation of the guided waves after having traveled through the bends. In the straight waveguide of Section\ref{ss:strip}, the waves within the structure are propagating almost perfectly parallel to the waveguide walls and hence they enter the absorber at a perpendicular angle. Such perpendicular incidence leads to the smallest possible reflections at an interface (as can be seen from the classical Fresnel equations, see, e.g., \cite{jackson1975electrodynamics}). In the Y-branch, after the waves pass through the junction and travel round the bends, it is to be expected that the waves will now have a traverse propagation component in addition to dominant longitudinal component. This means that the waves entering the absorbers appended to the right of the structure are doing so at a slightly oblique angle, leading to larger reflections. Therefore, when choosing appropriate length absorbers for bent waveguide structures, one must be cognisant of this effect. However, in the nanophotonics examples of interest here, where light is channeled by waveguides, the propagation direction is never too far from perfectly longitudinal, hence this effect will not lead to catastrophically large reflections from absorbers.

\subsection{Numerical aspects of the VIE method for photonics}
\label{ss:num_aspects}
To conclude the results section, we make some comments on the solution of our discretized integral equation. In particular, we discuss the solution via an iterative Krylov subspace method such as generalized minimum residuals (GMRES)~\cite{trefethen1997numerical}. First, we remind the reader that the main focus of this paper has been to solve the J-VIE (\ref{eqn:JVIE}) (we also considered the full JM-VIE (\ref{eqn:JM_VIE}) for interest, but for photonics applications only the J-VIE is required, see Section~\ref{sec:VIE}) which, when discretized, takes the form
\begin{equation}
	(\mat{I} - \mat{M}_{\epsilon} \mat{N})\matvec{w}_e = c_e\mat{M}_{\epsilon}\matvec{e}_{\text{inc}}.
	\label{eqn:mat_sys}
\end{equation}
For an iterative solver, we are only required to compute the matrix-vector product (MVP) of $\mat{I} - \mat{M}_{\epsilon} \mat{N}$ with a column vector. The discretized integral operator $\mat{N}$ is a \textit{block-Toeplitz} matrix with 6N unique entries, where $N$ is equal to the number of voxels for the piecewise constant basis function implementation of the VIE method employed here~\cite{polimeridis2014stable}. Therefore this dense operator only requires $\mathcal{O}(N)$ memory to be stored. Further, the MVP of $\mat{N}$ with a vector can be computed in $\mathcal{O}(N\log N)$ with the use of the FFT. Then, since $\mat{M}_{\epsilon}$ is diagonal, the total cost of the MVP with $\mat{I} - \mat{M}_{\epsilon} \mat{N}$ is also $\mathcal{O}(N\log N)$. Therefore the linear system (\ref{eqn:mat_sys}) can be solved via an iterative method such as GMRES with $\mathcal{O}(N\log N)$ cost.

A single MVP is fast owing to extremely efficient implementations of the FFT, e.g., \cite{frigo2005design}. However, if the matrix system (\ref{eqn:mat_sys}) is ill-conditioned or the eigenvalues of  $(\mat{I} - \mat{M}_{\epsilon} \mat{N})$ are not clustered near 1, then potentially hundreds or thousands of GMRES iterations, and hence MVPs, are required to solve the system. Thereby creating a huge constant in the aforementioned $\mathcal{O}(N\log N)$ cost. Thus keeping this iteration count small is crucial for the efficiency of the VIE method.

For low-frequency problems with the permittivity values considered here (for Si and SiO$_2$), only a handful of GMRES iterations are required to solve the integral equation, and so the VIE method employed without a precondtioner is extremely fast. However, as the number of wavelengths fitting across the domain of our problem increases, so does the iteration count. 

\begin{figure}[h!]
	\includegraphics[width=\linewidth]{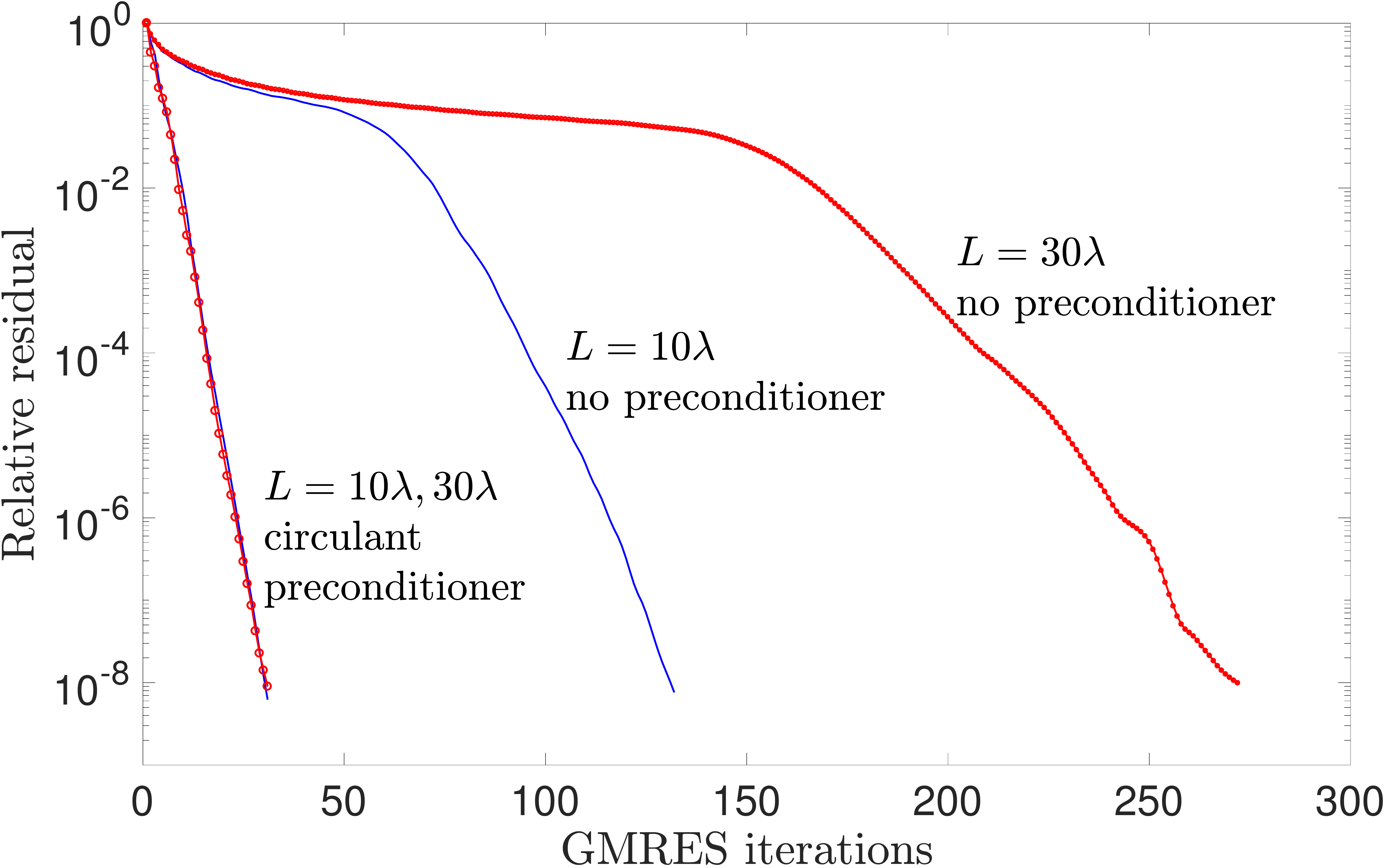}
	\caption{Convergence history of GMRES (tol $=10^{-8}$) without restarts for a straight waveguide of lengths $10\lambda$ and $30\lambda$. Observe how, with no preconditioner, the iteration count grows with the waveguide length. The circulant preconditioner, on the other hand, leads to an iteration count independent of the waveguide length.}
	\label{fig:GMRES_3}
\end{figure}

The problems arising in photonics involve light being channeled by waveguides such as that in Fig.~\ref{fig:strip}, such waveguides have typical dimensions
\[
	(X,Y,Z) \approx (20\lambda \rightarrow 2000\lambda, \lambda, \lambda/2),
\]
where $\lambda$ is the wavelength within the silicon. That is, the geometry is small in the $y$- and $z$-dimensions, but potentially very long (and hence high-frequency) in the $x$-dimension. For such high-frequency problems, the iteration count of GMRES is not small. In Fig.~\ref{fig:GMRES_3} we see the performance of GMRES with tolerance $10^{-8}$ and without preconditioner for the waveguide in Fig.~\ref{fig:strip} of lengths $10\lambda$ and $30\lambda$, and with a 4.5$\mu$m ($\approx 9\lambda$) quadratic absorber. Observe the stagnation of GMRES at a relative residual of around 0.1 before it rapidly converges; this is characteristic of high-frequency wave problems. It can be shown (we discuss this in more detail in a future publication) that this stagnation period, and hence the iteration count, increases approximately linearly with the waveguide length. This growth quickly leads to infeasibly large iteration counts. Therefore, one must seek to precondition the system (\ref{eqn:mat_sys}).

A popular preconditioning strategy for Toeplitz systems is to approximate the discrete operator by a circulant matrix. We employed such a strategy for all the computations in Sections~\ref{ss:strip} and \ref{ss:slow_light}. Here we give a brief overview of this strategy with in-depth details left to be provided in a separate article. We implemented a modified version of the technique proposed in \cite{chan1994circulant} to create a block-circulant matrix $\mat{W}$ which is closest to $(\mat{I} - \mat{M}_{\epsilon} \mat{N})$ in the Frobenius norm. Circulant matrices are diagonalized by the FFT, hence cheaply inverted. After constructing and inverting the circulant preconditioner, the following preconditioned system is solved via GMRES:
\begin{equation}
	\mat{W}^{-1}( \mat{I} - \mat{M}_{\epsilon}\mat{N})\matvec{w}_e= \mat{W}^{-1}c_e\mat{M}_{\epsilon}\matvec{w}_e.
\end{equation}
The preconditioned system has a matrix with eigenvalues well-clustered near unity and hence, as can be seen in Fig.~\ref{fig:GMRES_3}, the convergence of GMRES is greatly improved. Furthermore, we observe that the preconditioned iteration count is small and does not grow at all with the length of the waveguide. Therefore, the VIE method with circulant preconditioner is an extremely fast simulation tool for long nanophotonics structures.

\section{Conclusion}
\label{sec:conc}
Integral equation methods are traditionally used to simulate the scattering of waves from finite obstacles. However, when the obstacle is infinite in extent, such as nanophotonic waveguide structures, something must be done to truncate the domain in order to make the simulation feasible. In this paper, we presented and analyzed one such truncation approach, namely the introduction of adiabatic absorbing regions. The novelty of this paper lies in the application of adiabatic absorbers within the VIE method. In particular, we employ a VIE formulation that allows these absorbers to be introduced in a simple and straightforward manner which importantly does not affect the ``fast'' nature of the solver, thereby enabling rapid nanophotonics simulations.

We have outlined the application of the VIE method and the appropriate implementation of monomial adiabatic absorbers within the VIE setting. We have shown that the behavior of the reflections from these absorbers is in keeping with the theoretically obtained asymptotic results from coupled-mode theory. In particular, the transition reflections decay as $\mathcal{O}(L^{-2(d+1)})$, where $L$ is the length of the absorber and $d$ the degree of the monomial absorption profile. In Sections~\ref{ss:slow_light}, we performed simulations for a practical problem arising in photonics applications, namely the broadband simulation of the transmission through a Bragg grating. It was seen that, if the Bragg grating is simulated in its entirety and is truncated on the straight portions by adiabatic absorbers, accurate simulations result with short absorbers (approximately 5 wavelengths long). 

However, if the grating is truncated in the region of periodic modulation, extremely long absorbers are required to reduce transition reflections when near a band gap edge. Again, this is in keeping with asymptotic results from coupled-mode theory. In the nanophotonics applications of interest here, where devices are to be simulated in their entirety, one is unlikely to truncate a periodic structure with an absorber within the region of modulation. Nevertheless, adiabatic absorbers \textit{can} be applied, in contrast to perfectly matched layers which break down in such structures owing to the non-analyticity of the geometry~\cite{oskooi2008failure}. Therefore, with adiabatic absorbers in the VIE method, we can accurately simulate the propagation of waves within structures such as photonic crystals which have a large array of industrial applications~\cite{joannopoulos2011photonic}. That being said, there is a great deal of room for the optimization of adiabatic absorbers in these slow-light scenarios. Such optimization requires a careful study of the slow-light behavior and is sufficiently involved to warrant a separate article on the subject from the present authors.

The final structure we simulated was the Y-branch splitter. We saw that, with quadratic absorbers of length $5\lambda$, the reflections from the absorbers were negligible when compared to the total field. This was even in spite of the slightly larger than anticipated reflections from the right hand absorbers where the propagating waves had picked up a small transverse component due to traveling round the waveguide bends. When the waves enter the absorbers with off-perpendicular incidence, the reflections are increased. However, in nanophotonics structures, this transverse component will always be small, therefore, the increase in reflection will not be too large.

Finally, we presented some results pertaining to the iterative solution of the VIE's discrete system. For high-frequency problems (as encountered in photonics), the number of iterations required for an iterative solver to converge are large, regardless of the numerical method employed (e.g., finite difference, finite element, integral equation). Therefore, all numerical methods require effective preconditioners in order to make their application efficient. When the VIE (\ref{eqn:JVIE}) is discretized on a uniform grid, the resulting matrix in the discrete system has a three-level Toeplitz form. An effective preconditioner for this matrix can be obtained by making a circulant approximation on one or more levels of this Toeplitz matrix. The results presented in Section~\ref{ss:num_aspects} showed that such a preconditioner is extremely effective and renders the number of iterations small and independent of the structure's length. A more detailed study of this preconditioning strategy for VIEs shall also be presented in a separate article.

\section*{Acknowledgments}
This work was supported by a grant from Skoltech as part of the Skoltech-MIT Next Generation Program.

\bibliography{VIE_bib}

\begin{thebibliography}{10}
\providecommand{\url}[1]{#1}
\csname url@samestyle\endcsname
\providecommand{\newblock}{\relax}
\providecommand{\bibinfo}[2]{#2}
\providecommand{\BIBentrySTDinterwordspacing}{\spaceskip=0pt\relax}
\providecommand{\BIBentryALTinterwordstretchfactor}{4}
\providecommand{\BIBentryALTinterwordspacing}{\spaceskip=\fontdimen2\font plus
\BIBentryALTinterwordstretchfactor\fontdimen3\font minus
  \fontdimen4\font\relax}
\providecommand{\BIBforeignlanguage}[2]{{%
\expandafter\ifx\csname l@#1\endcsname\relax
\typeout{** WARNING: IEEEtran.bst: No hyphenation pattern has been}%
\typeout{** loaded for the language `#1'. Using the pattern for}%
\typeout{** the default language instead.}%
\else
\language=\csname l@#1\endcsname
\fi
#2}}
\providecommand{\BIBdecl}{\relax}
\BIBdecl

\bibitem{chrostowski2015silicon}
L.~Chrostowski and M.~Hochberg, \emph{Silicon photonics design: from devices to
  systems}.\hskip 1em plus 0.5em minus 0.4em\relax Cambridge University Press,
  2015.

\bibitem{lee2014low}
D.~H. Lee, S.~J. Choo, U.~Jung, K.~W. Lee, K.~W. Kim, and J.~H. Park,
  ``Low-loss silicon waveguides with sidewall roughness reduction using a
  {S}i{O}2 hard mask and fluorine-based dry etching,'' \emph{Journal of
  Micromechanics and Microengineering}, vol.~25, no.~1, p. 015003, 2014.

\bibitem{sarkar1986application}
T.~Sarkar, E.~Arvas, and S.~Rao, ``Application of {F}{F}{T} and the conjugate
  gradient method for the solution of electromagnetic radiation from
  electrically large and small conducting bodies,'' \emph{IEEE Transactions on
  Antennas and Propagation}, vol.~34, no.~5, pp. 635--640, 1986.

\bibitem{sertel2004multilevel}
K.~Sertel and J.~L. Volakis, ``Multilevel fast multipole method solution of
  volume integral equations using parametric geometry modeling,'' \emph{IEEE
  Transactions on Antennas and Propagation}, vol.~52, no.~7, pp. 1686--1692,
  2004.

\bibitem{chew2008integral}
W.~C. Chew, M.~S. Tong, and B.~Hu, ``Integral equation methods for
  electromagnetic and elastic waves,'' \emph{Synthesis Lectures on
  Computational Electromagnetics}, vol.~3, no.~1, pp. 1--241, 2008.

\bibitem{hanson2003green}
G.~Hanson, A.~Nosich, and E.~Kartchevski, ``Green's function expansions in
  dyadic root functions for shielded layered waveguides,'' \emph{Progress In
  Electromagnetics Research}, vol.~39, pp. 61--91, 2003.

\bibitem{kamandi2015integral}
M.~Kamandi, R.~Faraji-Dana \emph{et~al.}, ``Integral equation analysis of
  multilayered waveguide bends using complex images {G}reen's function
  technique,'' \emph{Journal of Lightwave Technology}, vol.~33, no.~9, pp.
  1774--1779, 2015.

\bibitem{oskooi2008failure}
A.~F. Oskooi, L.~Zhang, Y.~Avniel, and S.~G. Johnson, ``The failure of
  perfectly matched layers, and towards their redemption by adiabatic
  absorbers,'' \emph{Optics {E}xpress}, vol.~16, no.~15, pp. 11\,376--11\,392,
  2008.

\bibitem{alles2011perfectly}
E.~J. Alles and K.~W.~A. van Dongen, ``Perfectly matched layers for
  frequency-domain integral equation acoustic scattering problems,'' \emph{IEEE
  transactions on {U}ltrasonics, {F}erroelectrics, and {F}requency {C}ontrol},
  vol.~58, no.~5, pp. 1077--1086, 2011.

\bibitem{zhang2011novel}
L.~Zhang, J.~H. Lee, A.~Oskooi, A.~Hochman, J.~K. White, and S.~G. Johnson, ``A
  novel boundary element method using surface conductive absorbers for
  full-wave analysis of 3-d nanophotonics,'' \emph{Journal of Lightwave
  Technology}, vol.~29, no.~7, pp. 949--959, 2011.

\bibitem{polimeridis2014stable}
A.~Polimeridis, J.~F. Villena, L.~Daniel, and J.~White, ``Stable
  {F}{F}{T}-{J}{V}{I}{E} solvers for fast analysis of highly inhomogeneous
  dielectric objects,'' \emph{Journal of Computational Physics}, vol. 269, pp.
  280--296, 2014.

\bibitem{MARIE_GIT}
A.~Polimeridis and J.~F. Villena, ``{M}{A}getic {R}esonance {I}ntegral
  {E}quation ({M}{A}{R}{I}{E}) suite,''
  \url{https://github.com/thanospol/MARIE}, 2015.

\bibitem{povinelli2005slow}
M.~L. Povinelli, S.~G. Johnson, and J.~D. Joannopoulos, ``Slow-light, band-edge
  waveguides for tunable time delays,'' \emph{Optical Society of America},
  vol.~13, no.~18, pp. 7145--7159, 2005.

\bibitem{oughstun2003lorentz}
K.~E. Oughstun and N.~A. Cartwright, ``On the {L}orentz-{L}orenz formula and
  the {L}orentz model of dielectric dispersion,'' \emph{Optics {E}xpress},
  vol.~11, no.~13, pp. 1541--1546, 2003.

\bibitem{zhang2013compact}
Y.~Zhang, S.~Yang, A.~E.-J. Lim, G.-Q. Lo, C.~Galland, T.~Baehr-Jones, and
  M.~Hochberg, ``A compact and low loss {Y}-junction for submicron silicon
  waveguide,'' \emph{Optics {E}xpress}, vol.~21, no.~1, pp. 1310--1316, 2013.

\bibitem{volakis2012integral}
J.~Volakis, \emph{Integral equation methods for electromagnetics}.\hskip 1em
  plus 0.5em minus 0.4em\relax The Institution of Engineering and Technology,
  2012.

\bibitem{markkanen2012discretization}
J.~Markkanen, P.~Yla-Oijala, and A.~Sihvola, ``Discretization of volume
  integral equation formulations for extremely anisotropic materials,''
  \emph{IEEE Transactions on Antennas and Propagation}, vol.~60, no.~11, pp.
  5195--5202, 2012.

\bibitem{johnson2002adiabatic}
S.~G. Johnson, P.~Bienstman, M.~Skorobogatiy, M.~Ibanescu, E.~Lidorikis, and
  J.~Joannopoulos, ``Adiabatic theorem and continuous coupled-mode theory for
  efficient taper transitions in photonic crystals,'' \emph{Physical {R}eview
  E}, vol.~66, no.~6, p. 066608, 2002.

\bibitem{joannopoulos2011photonic}
J.~D. Joannopoulos, S.~G. Johnson, J.~N. Winn, and R.~D. Meade, \emph{Photonic
  crystals: molding the flow of light}.\hskip 1em plus 0.5em minus 0.4em\relax
  Princeton {U}niversity {P}ress, 2011.

\bibitem{jackson1975electrodynamics}
J.~D. Jackson, \emph{Electrodynamics}.\hskip 1em plus 0.5em minus 0.4em\relax
  Wiley Online Library, 1975.

\bibitem{trefethen1997numerical}
L.~N. Trefethen and D.~Bau~III, \emph{Numerical {L}inear {A}lgebra}.\hskip 1em
  plus 0.5em minus 0.4em\relax S{I}{A}{M}, 1997, vol.~50.

\bibitem{frigo2005design}
M.~Frigo and S.~G. Johnson, ``The design and implementation of {F}{F}{T}{W}3,''
  \emph{Proceedings of the IEEE}, vol.~93, no.~2, pp. 216--231, 2005.

\bibitem{chan1994circulant}
T.~F. Chan and J.~A. Olkin, ``Circulant preconditioners for {T}oeplitz-block
  matrices,'' \emph{Numerical Algorithms}, vol.~6, no.~1, pp. 89--101, 1994.

\end{thebibliography}
\bibliographystyle{IEEEtran}

\end{document}